\documentclass[review]{elsarticle}

\usepackage{lineno}
\usepackage{amsmath}
\usepackage{subfig}
\usepackage[toc,page]{appendix}
\usepackage{blindtext}
\usepackage{natbib}
\usepackage{siunitx}
\usepackage{relsize}
\usepackage{latexsym}
\usepackage{graphicx}
\usepackage{siunitx}
\usepackage{booktabs}
\usepackage{comment}
\usepackage{makecell}
\usepackage{relsize}
\usepackage{array,multirow,tabularx}
\usepackage{comment}

\usepackage[utf8]{inputenc}
\usepackage[T1]{fontenc}
\usepackage{lmodern}
\usepackage[figurename=Fig.,labelfont=bf,labelsep=period]{caption}
\usepackage{newtxtext,newtxmath}
\usepackage[citecolor=black,linkcolor=black]{hyperref}
\usepackage{tablefootnote}

\modulolinenumbers[5]

\journal{Extreme Mechanics Letters}

\biboptions{authoryear}

\begin{document}

\begin{frontmatter}

\title{
Physics-based Constitutive Modeling of Photo-oxidative Aging in Semi-Crystalline Polymers based on Chemical Characterization Techniques} 


\author{Aimane Najmeddine$^{1}$}
\author{Zhen Xu$^{2}$}
\author{Guoliang (Greg) Liu$^{2a}$}
\author{Alan R. Esker$^{2a}$}
\author{Maryam Shakiba\corref{mycorrespondingauthor}$^{1a}$}
\cortext[mycorrespondingauthor]{Corresponding author.}
\ead{mshakiba@vt.edu}
\address{$^{1}$Department of Civil and Environmental Engineering, Virginia Tech, USA}
\address{$^{2}$Department of Chemistry, Virginia Tech, USA}
\address{$^{a}$Macromolecules Innovation Institute, Virginia Tech, USA}

\begin{abstract}
This paper proposes a physio-chemically-based constitutive framework to simulate and predict the response of semi-crystalline low-density polyethylene (LDPE) to severe photo-oxidation. Photo-oxidation induced by exposure to Ultra-Violet (UV) light and oxygen is the dominant degradation mechanism affecting the lifespan of LDPE. In this work, we propose evolution functions for the material properties in the constitutive equations of \cite{boyce2000constitutive} to incorporate the effects of photo-oxidation on the mechanical response of LDPE. The evolution functions are based on chemically verified processes that are responsible for material degradation, namely the change in crystallinity and mass loss relative to the initial pristine films over exposure time. Changes in crystallinity and mass loss are characterized by Differential Scanning Calorimetry (DSC) and Quartz Crystal Microbalance with Dissipation Monitoring (QCM-D) experiments, respectively. Connecting the physio-chemical processes affecting polymer network evolution to the mechanical response of LDPE bypasses the need for defining fitting parameters that carry no physical meaning. The developed constitutive framework is validated with respect to a series of in-house uniaxial tensile tests performed on LDPE aged for different UV exposure times. Comparison of the constitutive framework versus experimental mechanical tests also confirms the accuracy of DSC and QCM-D as rigorous techniques to monitor and characterize degradation in LDPE films. The outcome shed light on the evolution of the macromolecular network in LDPE under extreme photo-oxidation and the evolution of the associated mechanical material properties. 


\end{abstract}


\end{frontmatter}



\section{Introduction}

Over the last few decades, semi-crystalline polymers have found their way into almost all outdoor structural applications (\textit{e.g.}, automotive and aerospace industries, electrical insulation technologies, and thermal storage applications) due to their excellent mechanical performance and optimal strength-to-weight ratio. During their service life, semi-crystalline polymers are exposed to several extreme environmental factors such as Ultra-Violet (UV) light, heat, oxygen, and other chemical processes that degenerate their mechanical properties and contribute to their permanent failure. In particular, UV light emitted by the sun or other artificial sources has been found to be the dominant degradation mechanism causing the fragmentation of semi-crystalline polymers into smaller-scaled particles known as microplastics \citep{yousif2013photodegradation,ranjan2019degradation,guo2019chemical}. Due to their minuscule sizes, microplastics can easily travel in large amounts through water pathways leading to the ocean. The abundance of microplastics in the marine environment has become a major concern in today's environmental discussion \citep{kershaw2015sources,brandon2016long,da2018degradation,bergmann2019white}. Therefore, it is imperative that special attention be devoted to the study of photo-oxidation impacts on semi-crystalline polymers for durable design and environmental preservation.


The presence of oxygen in addition to UV light accelerates polymer photodegradation and causes what is commonly referred to as photo-oxidation \citep{rabek1994polymer}. Generally, the resistance of polymers to photo-oxidation 
varies depending on the polymer composition, possible inherent contaminations, and the inclusion of pigments, additives, or fillers. Polymers with weak bond energies and high concentration of chromophoric groups (\textit{i.e.}, chemical groups that are capable of absorbing light) for instance, are generally more susceptible to photo-oxidation. %

Photo-oxidation and its deleterious effects on the lifespan of semi-crystalline polymers has been a subject of experimental investigation for decades. Photo-oxidation contributes to the degeneration of mechanical and aesthetic properties of semi-crystalline polymers and creates weaker materials that cannot sustain further mechanical loading, which ultimately leads to their complete failure \citep{CARRASCO20011457,HSU20122385,celina2013review,bhateja1983radiation,fayolle2008degradation,julienne2019influence,Hedir2020,cundiff2020photo}. In semi-crystalline polyolefins, for instance, photo-oxidation can be initiated either through hydroperoxide decomposition or through ketone photolysis via Norrish reactions \citep{rabek1994polymer}. As a result of these initiators, polymers can undergo an initial period of random chain-scission followed by a secondary period of crosslinking that is responsible for surface embrittlement. Due to this embrittlement, the polymers harden and visible cracks can potentially occur on their surface \citep{rodriguez2020effect}. A common consensus in the literature is that in semi-crystalline polymers, photo-oxidation reactions occur in the amorphous region that is favorable to oxygen diffusion \citep{ayoub2020modeling,rodriguez2020effect}. The random coil structure of the amorphous region favors chain linking/unlinking. As a result, when the polymer is exposed to light and oxygen, photo-oxidation-induced molecular chain alterations (\textit{i.e.}, chain-scission and crosslinking) manifest themselves in the unstructured, random amorphous phase. Therefore, given these considerations, it is clear that the macromolecular changes induced by photo-oxidation can be directly linked to the mechanical response (\textit{e.g.}, embrittlement, crack initiation and propagation, etc.) of photo-oxidatively aged polymers.




Many researchers have developed models to simulate the response of polymeric and elastomeric materials to environmental conditions (\textit{e.g.}, \cite{soares2008constitutive,soares2010deformation,vieira2014constitutive,vieira2011material,breche2016mechanical,breche2016non,wang2010entropy,han2009model,zhao2017oxidation,johlitz2014thermo,abdelaziz2019new,shakiba2021physics,XIAO201670,shakiba2016thermodynamic,ZHAO2020100826}). However, to the best knowledge of the authors, only a few studies tried to develop constitutive equations to study the behavior of semi-crystalline polymers in response to photo-oxidation. \cite{belbachir2010modelling} and \cite{ayoub2020modeling} used physics-based elasto-viscoplastic constitutive relationships to incorporate the effects of UV radiation on the mechanical properties of polylactic acid (PLA) and low-density-polyethylene (LDPE), respectively. More recently, \cite{lamnii2021experimental} captured the effect of UV radiation on the fatigue life of a bulk semi-crystalline polymer based on two fatigue indicators: the maximum true stress and the dissipated energy. The authors used the evolution of the molecular weight of photo-oxidatively aged polymers to define a degradation parameter suitable for macromechanical response prediction. However, a major limitation to all these studies 
concerned the identification of the evolution of material properties 
which contained fitting parameters that carried no physical meaning. The evolution of the material properties in these studies was obtained simply by fitting the constitutive equations to the already obtained experimental mechanical measurements on aged samples. Doing so renders the constitutive equations essentially a fitting algorithm that can only describe the particular scenario upon which calibration was performed. In contrast, purely physio-chemically-based evolution functions of the material properties, based on network evolution, are desirable to eliminate the need for fitting parameters. 

Thus, although works have been accomplished in the experimental and the numerical sides, the link between the network evolution and the mechanical responses in photo-oxidative aging of polymers is still missing.  
In this study, we investigate the effects of photo-oxidation on the mechanical performance of LDPE. 
Our goal is to present a physio-chemically-based constitutive framework to predict the macromechanical behavior of LDPE materials in response to photo-oxidation. More severe UV radiation, compared to the previous studies, is also considered in this work.



This manuscript is organized as follows. Section~\ref{objectives} reiterates the objectives and main contributions of this work. Section~\ref{constitutive model} provides a concise description of the constitutive framework that has been adopted in this work to describe the mechanical response of unaged LDPE. In section~\ref{photo_kinetics}, a detailed discussion on photo-oxidation processes is provided to propose a novel methodology to monitor changes in the material properties of LDPE due to photo-oxidation. Then, upon identification of the physical processes responsible for photo-oxidation of LDPE, we present in section~\ref{material_charac} the in-house experimental investigations proposed to determine our evolution functions for the material properties. Results and their corresponding analyses are provided in section~\ref{results}. Finally, section~\ref{conclusion} concludes with some important remarks and ideas for subsequent future investigations. 

\section{Objectives} \label{objectives}

This paper will contribute to the missing relationship between the chemical macromolecular changes and the mechanical responses of semi-crystalline LDPE due to photo-oxidation. To achieve this, a framework is developed to connect the evolution of the material properties in the constitutive equations to the physio-chemical processes affecting polymer network. 
This connection eliminates the need to conduct mechanical testing on aged polymers and bypasses the need for extra fitting parameters. Our objectives are summarized as follows:
\begin{itemize}
    \item First, based on our understanding of how photo-oxidation affects semi-crystalline polymers, we aim to develop a physically-based and chemically-motivated constitutive framework to predict the response of photo-oxidatively aged LDPE. 
    
    The chemical characterization techniques employed in this work are Differential Scanning Calorimetry (DSC) and a Quartz Crystal Microbalance with Dissipation Monitoring (QCM-D). DSC is used to determine the evolution of the crystallinity, whereas QCM-D is used to determine the evolution of the minute mass ratio between the initial unaged thin polymer and the corresponding aged samples.
    \item Second, we plan to verify the validity of employing the above characterization techniques (particularly QCM-D) on thin polymers to investigate the photo-oxidation of relatively thicker films. 
\end{itemize}

Although the mass loss characterizes the direct damage from physio-chemical reactions of the polymer during aging, to the best of our knowledge, there exists currently no study that uses the mass loss evolution to quantify the degree of photo-oxidation in polymers. The primary reason of the missing of mass ratio as a damage indicator is the difficulty of mass change determinations at microgram or even nanogram scale. Minute mass determination based on ordinary techniques is often unreliable due to the low sensitivity of analytical balance to mass change at micro- or nanoscale. In contrast, QCM-D, an acoustic technique, provides sensitive detection of mass change and high accuracy to nanogram-scale, which can be a solution to this challenge, thus fulfilling the deficiency of minute mass loss evolution in the literature.

\section{Constitutive and Governing Equations} \label{constitutive model}

Our goal is to predict the macromechanical response of photo-oxidatively aged LDPE based solely on the fundamental understandings of the chemical macromolecular changes occurring in the material upon exposure to UV radiation. The effect of photo-oxidation is captured by conjecturing appropriate chemistry-based evolution functions for the mechanical properties of LDPE. We begin with describing the constitutive relationships governing the mechanical response of semi-crystalline polymers, and later present the framework accounting for the contribution of photo-oxidation to polymer mechanical degradation.  

\begin{figure*}[h!bt]
    \centering
    \subfloat[]{
        \includegraphics[width=0.48\linewidth]{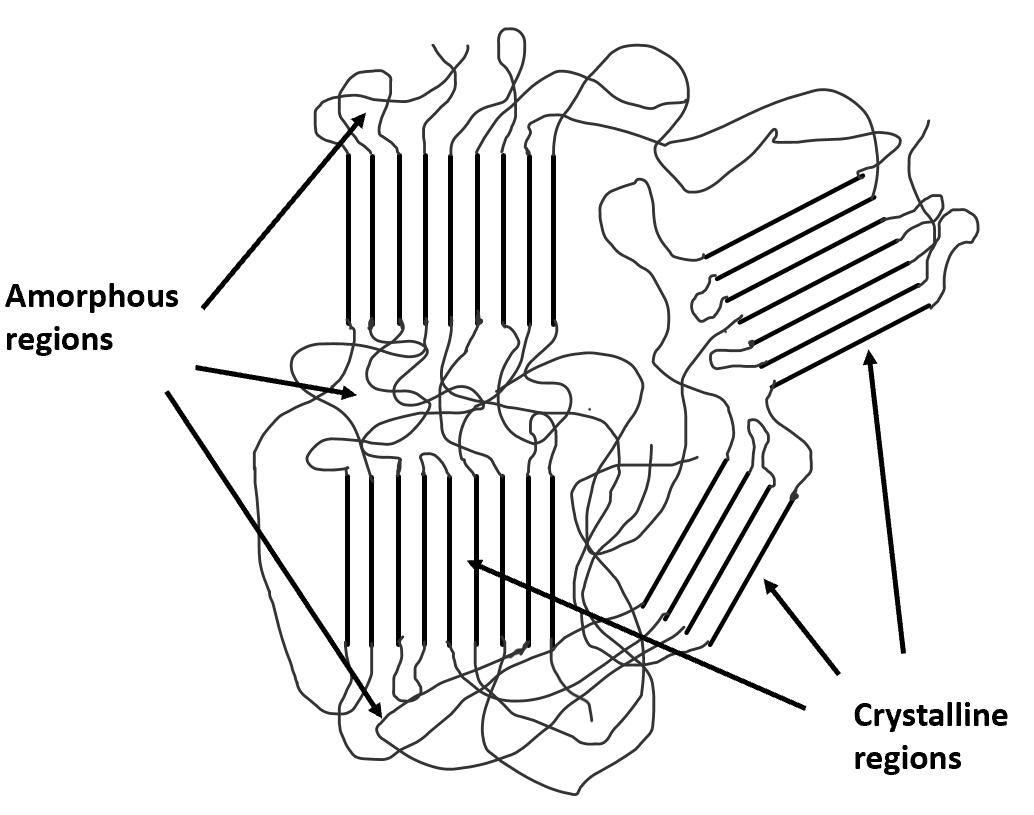}
        \label{SM}
        }
        \hfill
    \subfloat[]{
        \includegraphics[width=0.48\linewidth]{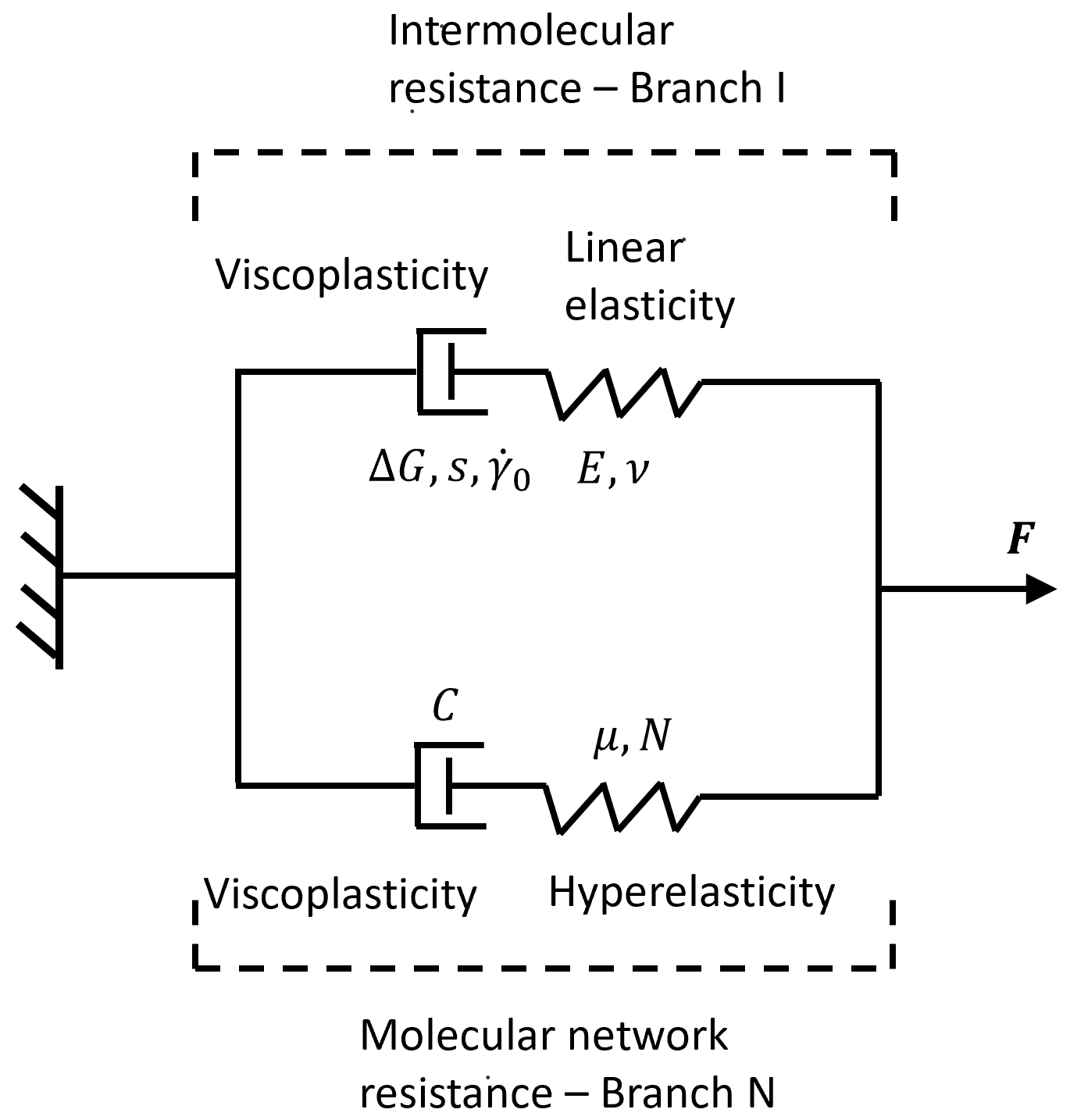}
        \label{RM}
        }        
    \caption{a) Schematic representation of a semi-crystalline polymer consisting of two contributing regions: an amorphous region characterized by the coil-like random arrangement, and a crystalline region characterized by the structured, orderly geometric alignment. b) Rheological representation of the constitutive theory of \cite{boyce2000constitutive} highlighting the contribution from the intermolecular resistance (\textit{i.e.}, branch I) and molecular network resistance (\textit{i.e.}, branch N). The elasto-viscoplastic parameters are shown attached to each corresponding element. The effects of both the amorphous as well as the the crystalline regions on the mechanical resistance to deformation is captured implicitly via the elastic modulus $E$. The total deformation gradient $\mathbf{F}$ is applied to both branches. } 
    \label{fig: Boyce-Socrate-Llana}
\end{figure*}

A number of studies in the literature have tried to develop constitutive relationships to describe the finite-strain elasto-viscoplastic behavior of polymers \citep{boyce1988large,arruda1995effects,bardenhagen1997three,tervoort1997constitutive,boyce2000constitutive,ahzi2003modeling,anand2003theory,makradi2005two,dupaix2006constitutive,ayoub2010modelling}. The three-dimensional physics-based constitutive theory of \cite{boyce2000constitutive} -- which is what we used in this work -- is particularly attractive due to its simplicity and its capability of simulating various behaviors of thermoplastics based on the motion of molecular chains. The constitutive relationships of \cite{boyce2000constitutive} were originally developed to describe deformation resistance of amorphous polymers processed above their glass transition temperature; however, in semi-crystalline polymers, the contribution due to crystallinity shown in Figure~\ref{SM} can also be captured implicitly through the elastic modulus \citep{abdul2014large}. 

In the constitutive formulation of \cite{boyce2000constitutive}, the resistance to deformation consists schematically of two nonlinear Maxwell elements connected in parallel to one another as shown in Figure~\ref{RM}. Branch I involves a linear elastic spring to represent molecular interactions, and a nonlinear viscous dashpot to account for the non-Newtonian flow arising from the motion of polymer segments (unlinking and sliding) as shown in Figure~\ref{SM}. The spring stiffness in Branch I implicitly considers the contribution from both the amorphous as well as the crystalline phases illustrated in Figure~\ref{SM}. Branch N is composed of a nonlinear elastic spring (\textit{i.e.}, Langevin spring) representing the rubbery behavior of the polymer network based on the non-Gaussian statistical mechanics theory of rubber elasticity \citep{arruda1993three}. The nonlinear spring is intended to capture the post-yield strain hardening at large strains due to the alignment of the long-chain polymer molecules. A nonlinear dashpot that is connected in series to it is included to represent the rate- and temperature-dependent flow arising from the motion of polymer segments at large strains. The inclusion of the two nonlinear dashpots captures the rate-dependency of the stress-strain behavior through molecular orientation and relaxation. 

Since the branches of the schematic representation shown in Figure~\ref{RM} are parallel, the total deformation gradient $\mathbf{F}$ is applied to both branches and we have:
\begin{align} \label{parallel def}
    \mathbf{F} = \mathbf{F}_I = \mathbf{F}_N
\end{align}
where the indices I and N refer to the intermolecular and network resistance branches, respectively. The total Cauchy stress tensor $\mathbf{T}$ can therefore be written as the sum of the two contributions $\mathbf{T}_I$ and $\mathbf{T}_N$:
\begin{align}
    \mathbf{T} = \mathbf{T}_I + \mathbf{T}_N
\end{align}

Next, we present details on the kinematic configuration as well as the developed constitutive framework considering photo-oxidation effects.



\subsection{Kinematics}

Both branches involve springs that are attached in series to dashpots. Therefore, the deformation gradients corresponding to each branch can be decomposed multiplicatively into elastic and plastic components as follows \citep{lee1969elastic}: 
\begin{align} \label{lee decomp}
    \mathbf{F}_I = \mathbf{F}^e_I \mathbf{F}^p_I
    \quad\text{and}\quad
    \mathbf{F}_N = \mathbf{F}^e_N \mathbf{F}^p_N
\end{align}
where the superscripts $e$ and $p$ refer to the elastic and plastic parts, respectively. 

The plastic deformation gradients $\mathbf{F}^p_I$ and $\mathbf{F}^p_N$ can be obtained as follows:
\begin{align} \label{dot FIp}
    \dot{\mathbf{F}}_I^p = {\mathbf{F}_I^e}^{-1} \mathbf{D}_I^p
    \mathbf{F}_I^e \mathbf{F}_I^p
    \quad\text{and}\quad
    \dot{\mathbf{F}}_N^p = {\mathbf{F}_N^e}^{-1} \mathbf{D}_N^p
    \mathbf{F}_N^e \mathbf{F}_N^p
\end{align}
in which the rates of inelastic deformation $\mathbf{D}_I^p$ and $\mathbf{D}_N^p$ must be described. The dot expression denotes the time derivative $\frac{\partial}{\partial t} (\cdot)$. The elastic deformation gradients $\mathbf{F}_I^e$ and $\mathbf{F}_N^e$ are obtained using Equations~\ref{lee decomp}. The derivations leading to Equations~\ref{dot FIp} are provided in Appendix A.   

\subsection{Intermolecular contribution}

In this section, we explain the governing visco-elastoplastic equations corresponding to the intermolecular resistance branch.  
The intermolecular resistance is represented by a linear spring in series with a nonlinear dashpot. The intermolecular Cauchy stress $\mathbf{T}_I$ is expressed in terms of the Hencky strain ln$(\mathbf{V}_I^e)$ as:
\begin{align} \label{eq: intermolecular Cauchy Stress}
    \mathbf{T}_I = \frac{1}{J_I^e} \mathbf{C}_I^e \text{ln}(\mathbf{V}_I^e)
\end{align}
where $J_\text{I}^e = \text{det} \mathbf{F}_I^e$ and $\mathbf{C}_I^e$ is the isotropic fourth-order elastic stiffness tensor expressed as: 
\begin{align} \label{eq: fourth-order elastic stiffness tensor}
    \mathbf{C}_I^e = \Big (\frac{E}{1+\nu} \Big) \mathbf{\mathbb{I}} + \Big (\frac{E \nu}{(\nu + 1)(2\nu -1 )} \Big) \mathbf{\mathcal{I}} \otimes \mathbf{\mathcal{I}}
\end{align}
where $E$ is the Young's modulus, $\nu$ is the Poisson's ratio, and $\mathbf{\mathbb{I}}$ and $\mathbf{\mathcal{I}}$ are the fourth- and second-order identity tensors, respectively. The symbol $\otimes$ denotes the tensor product operation. 

The viscoplastic strain rate tensor $\mathbf{D}^p_I$ is expressed by the following flow rule:
\begin{align} \label{eq: viscoplastic flow rule}
    \mathbf{D}^p_I = \dot\gamma^p_I \frac{\textbf{DEV}(\mathbf{T}_I)}{\sqrt{2}\tau_I}
\end{align}
where $\textbf{DEV}(\mathbf{T}_I) = \mathbf{T}_I - \frac{tr(\mathbf{T}_I)}{3} \mathbf{\mathcal{I}} $ is the deviatoric part of $\mathbf{T}_I$, $\tau_I = \frac{1}{\sqrt{2}} ||\textbf{DEV}(\mathbf{T}_I)||$ 
is the effective shear stress written in terms of the Frobenius norm $||\textbf{DEV}(\mathbf{T}_I)||$ of the deviatoric part of $\mathbf{T}_I$, and $\dot\gamma^P_I$ is the viscoplastic shear strain rate given by the following expression:
\begin{align} \label{eq: viscoplastic shear strain}
    \dot\gamma^p_I = \dot\gamma_0 exp\Big[-\frac{\Delta G_a}{K_B\Theta}\bigg(1-\frac{\tau_I}{s}\Big)\bigg]
\end{align}
where $\dot\gamma_0$, $\Delta G_a$, $s$, $K_B$, and $\Theta$ are the pre-exponential factor, the activation energy, the athermal shear strength, the Boltzmann constant, and the absolute temperature, respectively.

\subsection{Network contribution}

In this section, we explain the governing visco-hyper-elastoplastic equations corresponding to the molecular network resistance branch N. The molecular network resistance is represented by a nonlinear spring in series with a nonlinear dashpot. 
The molecular network part of the Cauchy stress $\mathbf{T}_N$ is expressed as a function of the elastic deformation gradient $\mathbf{F}_N^e$ using a non-Gaussian statistical framework involving the inverse of the Langevin function $\mathcal{L}^{-1}$ \citep{arruda1993three}:
\begin{align} \label{eq: Arruda-Boyce stress tensor} 
\mathbf{T}_N = \frac{1}{J^e_N} \mu \frac{\sqrt{N_k}}{\bar \lambda^e_N} \mathcal{L}^{-1}\bigg(\frac{\bar\lambda^e_N}{\sqrt{N_k}}\bigg) \Big[ \mathbf{\bar B}_N^e - (\bar\lambda^e_N)^2 \mathbf{\mathcal{I}}    \Big] 
\end{align}
where $\mu = n K_B \Theta$ is the rubber modulus given as a function of $n$ the number of chains per unit volume, $N_k$ is the number of Kuhn monomers per chain, $\mathcal{L}(\cdot) = \text{coth}(\cdot) - \frac{1}{(\cdot)}$ is the Langevin function whose inverse is given by $\mathcal{L}^{-1}(x) = x \bigg( \frac{3-x^2}{1-x^2} \bigg)$ \citep{cohen1991pade}, and $\bar\lambda^e_N = \sqrt{\frac{\bar I_1}{3}}$ is the effective macro-stretch written as a function of the first invariant $\bar I_1=tr(\mathbf{\bar B}^e_N)$ of the elastic isochoric left Cauchy-Green deformation tensor $\mathbf{\bar B}^e_N = (J^e_N)^{-2/3}\mathbf{F}^e_N{\mathbf{F}^e_N}^T$.


The flow strain rate tensor $\mathbf{D}^p_N$ is expressed by the following flow rule:
\begin{align} \label{eq: flow strain rate}
    \mathbf{D}^p_N = \dot\gamma^p_N \bigg (\frac{\textbf{DEV}(\mathbf{T}_N)}{\sqrt{2}\tau_N} \bigg)
\end{align}
where $\textbf{DEV}(\mathbf{T}_N) = \mathbf{T}_N - \frac{tr(\mathbf{T}_N)}{3} \mathbf{\mathcal{I}} $ is the deviatoric part of $\mathbf{T}_N$, $\tau_N = \frac{1}{\sqrt{2}} ||\textbf{DEV}(\mathbf{T}_N)||$
is the effective shear stress, and $\dot\gamma^p_N$ is the flow shear strain rate given by the following expression:
\begin{align} \label{eq: flow shear strain rate}
    \dot\gamma^p_N = C \bigg( \frac{\tau_N}{\lambda^p_N - 1} \bigg)
\end{align}
in which the parameter $C$ is included to account for temperature-dependency of relaxation and $\lambda^p_N = [tr(\mathbf{B}_N^p)/3]^{1/2}$, where $\mathbf{B}_N^p = \mathbf{F}_N^p{\mathbf{F}^p_N}^T$. Note that Equation~\ref{eq: flow shear strain rate} is unstable for $\lambda^p_N = 1$; therefore, to ensure numerical stability, a perturbation coefficient equal to $10^{-6}$ is added to  $\lambda^p_N$ in all of our simulations.

\subsection{Photo-oxidation contribution} \label{photo_kinetics}

\begin{figure*}[hbt!]
    \centering
    \includegraphics[scale=0.4]{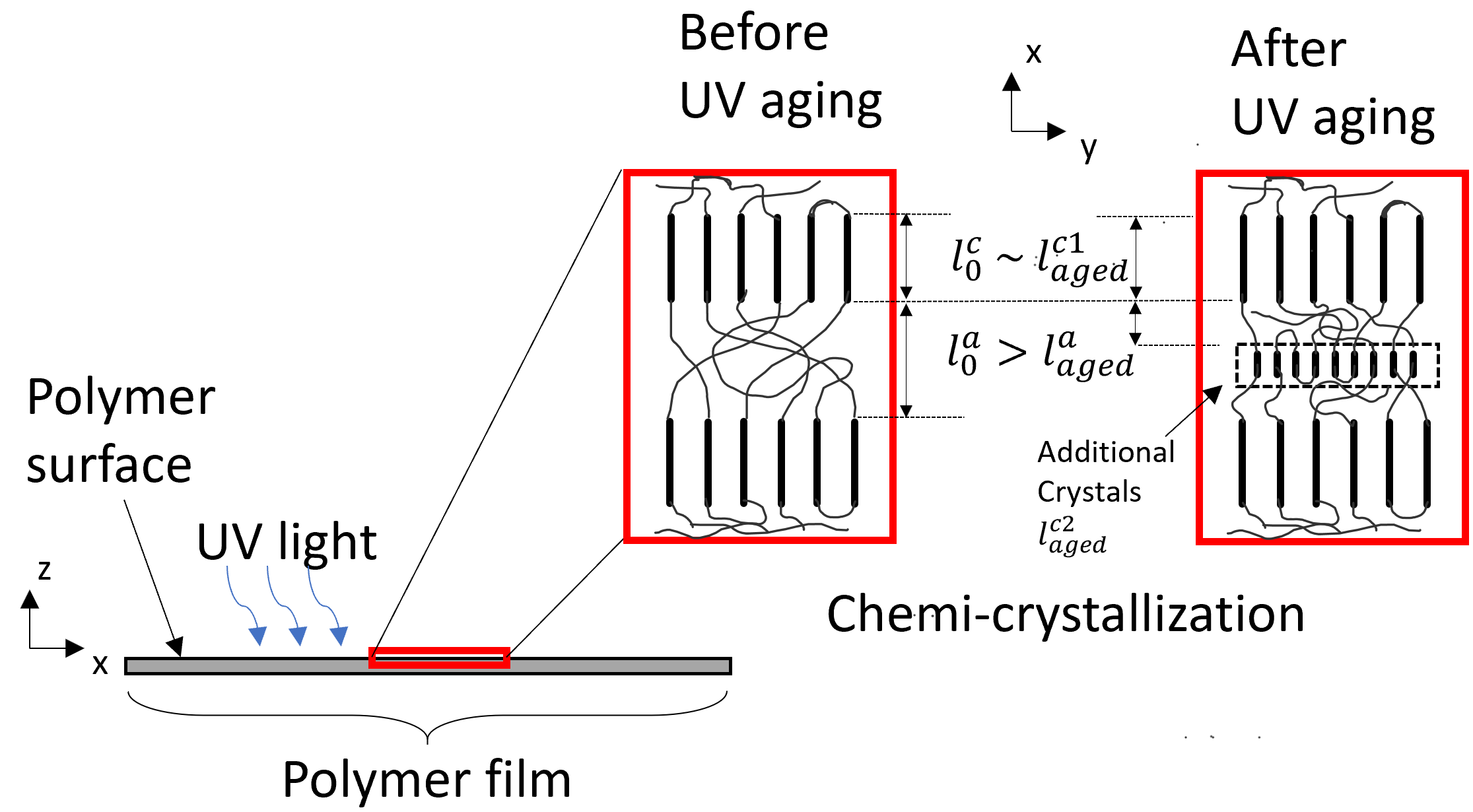} 
    \caption{Schematic representation of chemi-crystallization due to photo-oxidation. Upon exposure to UV light, the molecular chains in the amorphous region break and degrade causing the formation of additional crystals within the amorphous domain. Thickness of the primary crystalline region remains unchanged but the thickness of the amorphous domain decreases, thus allowing free segments to re-crystallize within a newly formed crystalline domain. In other words, the amorphous region shrinks ($ l^a_{aged} < l^a_0$) on the expense of the crystalline phase that gains further structuring ($ l^c_0 < l^{c1}_{aged} + l^{c2}_{aged} $). 
    }
    \label{degradation_mechanism}
\end{figure*}

Photo-oxidation induces alterations to the mechanical properties of semi-crystalline polymers. Upon exposure to UV light, semi-crystalline polymers undergo an initial period of chain-scission in which long molecular chains in the amorphous phase break causing a decrease in the average molar mass. As a result, segments of entangled chains in the amorphous region are released, and with enough mobility, free segments can rearrange into a crystalline region \citep{rabello1997crystallization}. With increased crystallinity, the inter-lamellar spacing decreases and embrittlement takes place \citep{fayolle2008degradation}. This process is known as chemi-crystallization and is illustrated schematically in Figure~\ref{degradation_mechanism}. 
During photo-oxidation, thickness of the primary crystalline region remains unchanged ($ l^c_0 \approx l^{c1}_{aged}$); however, thickness of the amorphous domain decreases ($ l^a_{aged} < l^a_0$) and free segments begin to re-crystallize within a newly formed crystalline domain of thickness $l^{c2}_{aged}$ \citep{rodriguez2020effect}. In other words, the amorphous region shrinks at the expense of the crystalline phase that gains further structuring ($ l^c_0 < l^{c1}_{aged} + l^{c2}_{aged} $). Note that we keep the distinction between primary and secondary crystallites out of discussion since such distinction is irrelevant from the perspective of the behavior of the material where crystallinity is expected to increase regardless of which label is most appropriate.

On the one hand, chemi-crystallization is indicative of a stiffening and a strengthening behavior probing intermolecular interactions. In fact, increase in crystallinity (and consequent shrinkage of the amorphous domain) as shown in Figure~\ref{degradation_mechanism} suggests that an increased stiffness is expected. On the other hand, chemi-crystallization also indicates an increase in the flow stress required to overcome intermolecular barriers to deformation. Therefore, it is expected that the material properties involved in the intermolecular resistance branch to change in response to photo-oxidation.  

In this section, we aim to conjecture appropriate evolution functions for the material properties corresponding to the intermolecular resistance branch, \textit{i.e.}, the initial stiffness $E$, the athermal shear strength $s$, and the activation energy $\Delta G_a$, based on the chemical understanding presented heretofore.  

\textbf{\textit{Remark}}. Recall that the material parameters involved in the constitutive relationships of \cite{boyce2000constitutive} are: for the intermolecular resistance branch, the elastic modulus $E$, the Poisson's ratio $\nu$, the athermal shear strength $s$, the activation energy $\Delta G_a$, and the pre-exponential factor $\dot \gamma_0$, whereas for the network resistance branch, the rubber modulus $\mu$, the number of Kuhn monomers per chain $N_k$, and the constant accounting for temperature-dependency of relaxation $C$. The parameters $\nu$, $\dot \gamma_0$, and $C$ are assumed constant in this work. The reason for this assumption is provided as follows. Since LDPE films are tested above their glass transition temperature, we assign the value $0.49$ to the Poisson's ratio $\nu$. Additionally, at room temperature -- which is where accelerated photo-oxidation aging is performed in this study -- the temperature-dependent relaxation parameter $C$ is assigned the value $8 \times 10^{-8}$ $\text{MPa}^{-1}$ \citep{boyce2000constitutive}. Finally, the pre-exponential factor $\dot\gamma_0$ is assigned the value $1.75\times10^6$ $\text{s}^{-1}$ \citep{boyce2000constitutive}. On the other hand, the rubber modulus $\mu$ and the number of monomers per chain $N_k$ are assumed to have minor effects on the response of LDPE to photo-oxidation, especially at long aging times. In fact, 
as the material becomes highly crystalline, it also becomes brittle and fracture occurs prematurely when mechanical load is applied (\textit{i.e.}, at lower strain levels). The increase in crystallinity and subsequently the premature fracture of LDPE at long aging times means that the resistance to deformation at long aging times can be captured simply by the intermolecular branch responsible for the elasto-plastic behavior. Therefore, from a mathematical standpoint, the material properties corresponding to the network resistance branch (\textit{i.e.}, the rubber modulus $\mu$ and the number of Kuhn monomers per chain $N_k$) -- which govern the large-strain deformation behavior, are no longer necessary. Instead, at long exposure times, the material is highly brittle and the mechanical response can be captured elasto-plastically (\textit{i.e.}, using only the intermolecular branch contribution). 

In order to conjecture appropriate chemistry-based evolution expressions for the material properties, we first recognize that photo-oxidation effects are mostly surface effects \citep{suresh2011mechanical,shlyapintokh1983synergistic,yousif2013photodegradation}. Photo-oxidation causes carbonyl groups to form on the surface of polyethylene and increases hydrophilicity, which then leads to embrittlement \citep{suresh2011mechanical}. As a result, photo-oxidation-induced damage concentrates on the surface of the material. Taking these considerations into account, we propose to employ the minute mass ratio, utilizing a surface-sensitive technique, as the characteristic degradation indicator to photo-oxidation in LDPE. Additionally, as mentioned in the earlier discussion, changes in the molecular-surface interactions as well as the crystallinity probe the evolution of the initial stiffness in response to photo-oxidation. Therefore, we posit that the evolution of the Young's modulus where the material is degraded be expressed as follows:
\begin{align} \label{eq: evolution of E} 
 E(t) = E_{0} \bigg(\frac{\zeta(t)}{\zeta_{0}}\bigg) \omega(t)^{-1}
\end{align}
where $E_{0}$ is the Young's modulus at the initial pristine configuration, $\zeta_{0}$ and $\zeta(t)$ are the crystallinities at the initial (unaged) and current (aged for some aging time, t) configurations, respectively, and $\omega(t)$ is the degradation indicator defined as minute mass ratio between the current and initial aging states (\textit{i.e.}, $\omega(t)=m(t)/m(0)$ where $m(0)$ and $m(t)$ are the polymer masses at UV exposure duration of 0 and t, respectively). 

Equation~\ref{eq: evolution of E} captures two principles. First, the initial stiffness depends on the evolution of the crystallinity. In fact, not only does this dependence capture the increase in the initial modulus, but even situations for which the initial stiffness decreases or remains constant can be well captured. Indeed, the crystallinity can follow any type of evolution depending on the chemical mechanism at hand. The stiffness will then follow a similar evolution (with mass loss held fixed) due to the linear proportionality of Equation~\ref{eq: evolution of E}. Second, the effect of mass degeneration on the stiffness is taken into account through an inverse proportionality. This inverse dependence can be justified as follows. First, as previously mentioned in the manuscript, photo-oxidation effects occur predominantly in the amorphous region of semi-crystalline polymers. Therefore, any changes in the polymer's mass would suggest that the amorphous phase is perturbed on the expense of the crystalline region (see Figure~\ref{degradation_mechanism}). Additionally, one could appreciate the parallelism between Equation~\ref{eq: evolution of E} and the relationship for elastomers relating the crosslink density, $\rho$, and the molar mass between two crosslinks, $M_c$, through $\rho \propto \frac{1}{M_c}$. It suffices to mention that the crosslink density in elastomers is proportional to stiffness to deduce the inverse proportionality between molar mass and stiffness.  

In Equation~\ref{eq: evolution of E}, aside from $E_{0}$ which can be obtained directly from a mechanical tensile test on an unaged sample, the evolution of the Young's modulus is given entirely as a function of variables that can be experimentally determined through appropriate chemical characterization tests -- in this case, DSC to obtain the crystallinity $\zeta$ and the surface-sensitive technique QCM-D to obtain the degradation parameter $\omega$. Therefore, no additional fitting variables are required, making Equation~\ref{eq: evolution of E} purely physio-chemically motivated. Therefore, in defining Equation~\ref{eq: evolution of E}, we have imparted actual physical meaning to the micromechanical changes of the polymer instead of simply assuming the usual empirical approach that results in numerous extra fitting parameters bearing no actual physical meaning.

The effect of photo-oxidation on the evolution of the remaining physical variables on the intermolecular branch of Figure~\ref{RM} (\textit{i.e.}, the athermal shear strength $s$ and the activation energy $\Delta G_a$ in Equation~\ref{eq: viscoplastic shear strain}) is accounted for by describing appropriate evolution functions in terms of the degradation indicator $\omega(t)$. In particular, the evolution of the athermal shear strength $s$ is defined as follows: 
\begin{align} \label{eq: evolution of s}
 s(t) = s_{0} \omega(t)^{-1}
\end{align} 
where $s_0$ and $s(t)$ are the athermal shear strengths corresponding to the unaged and aged states, respectively. Again, the inverse proportionality is justified using the same argument as the one given for the Young's modulus. Furthermore, assuming that the ratio $\frac{\Delta G_a}{s}$ remains constant \citep{belbachir2010modelling}, the evolution of the activation energy can be expressed as follows: 
\begin{align} \label{eq: evolution of delG}
 \Delta G_a(t) = \Delta G_{a0} \bigg(\frac{s(t)}{s_{0}} \bigg)
\end{align} 
where $\Delta G_{a0}$ is the activation energy corresponding to the initial unaged state. 

Having identified the responsible mechanisms for degradation and conjectured appropriate physio-chemically based evolution functions for the material properties, in section~\ref{material_charac}, we present the chemical characterization techniques that have been designed to measure the crystallinity and the minute mass ratio needed in the evolution equations~\ref{eq: evolution of E}, ~\ref{eq: evolution of s}, and ~\ref{eq: evolution of delG}. 

\section{Material Characterization} \label{material_charac}

In this section, information pertaining to the experimental techniques is summarized. Specific details regarding the LDPE material, the UV aging procedure for which the LDPE films were subjected to, and the experimental test measurements (\textit{i.e.}, DSC, QCM-D, and quasi-static tensile tests) are presented. Interpretation of the experimental results is provided in section~\ref{exp_results_interpretation}. 

\subsection{Material}

LDPE pellets were purchased from Sigma-Aldrich and used as received. The density of the virgin LDPE is 0.93 $g/cm^2$ and the melting point is $\SI{116} {\celsius}$. LDPE films were prepared from thermopressing at $\SI{180} {\celsius}$ with loading of 8 $tons$ during 2 $min$. The resulting films were cooled in air from $\SI{180} {\celsius}$ to room temperature and were subsequently thermally annealed at $\SI{110} {\celsius}$ for 1 $h$. The resulting polymer films had thicknesses ranging between 30 and 80 $\mu m$. This range of thickness was intentionally selected to allow for homogeneous oxidation and prevent diffusion-limited-oxidation (DLO) conditions \citep{ayoub2020modeling,tavares2003effect,tireau2009environmental,hsueh2020micro}.

\subsection{UV aging} \label{aging_procedure}
Polymer films were aged under a 250 $W$ UV lamp at a wavelength of 254 $nm$ (Rayonet with a maximum UV dose of 125 $kW/m^2$) to simulate and accelerate the LDPE photo-oxidation in air at room temperature ($\SI{25} {\celsius}$). Subsequently, the LDPE coated QCM-D sensors were aged in the UV chamber for varying aging times (\textit{i.e.}, 0, 24, 48, 72, and 112 $h$). It is worth comparing the above UV dose to solar radiation which has a UV intensity of approximately 100-200 $W/m^2$ , the maximum aging time chosen in this work (112 $h$) corresponds to 432 days of solar radiation. 


\subsection{Differential Scanning Calorimetry}

DSC was performed between 40 and $\SI{200} {\celsius}$ at a heating rate of $\SI{30} {\celsius}/min$ under a Nitrogen stream of 50 $mL/min$ on a Discovery DSC2500 (TA Instruments). The DSC was calibrated using indium (melting point (m.p.) = $\SI{156.60} {\celsius}$) and zinc (m.p. = $\SI{419.47} {\celsius}$) standards. The crystallinity of LDPE was calculated using melting enthalpy divided by 293 $J/g$ for 100\% crystalline material. DSC characterization on films with thicknesses smaller than 1 $\mu m$ is difficult due to the low sensitivity of common DSC at low sample mass. Therefore, crystallinity was measured for bulk samples (\textit{i.e.}, samples with thicknesses ranging between 30 and 80 $\mu m$). This range of thickness is larger than the threshold limit for which thickness effects on crystallinity measurement become significant, \textit{i.e.}, between 300 $nm$ and 1000 $nm$ \citep{wang2004crystallization}. 

\subsection{Quartz Crystal Microbalance with Dissipation Monitoring}

QCM-D is a mass measurement technique that is highly surface sensitive and is mostly employed to measure the mass of layers in the nanometer thickness range. In this study, QCM-D was utilized to measure the minute mass ratio between the aged and unaged films. To investigate the effect of film thickness on the mass loss, three varying thicknesses (three parallel samples each) were prepared (\textit{i.e.}, 146, 158, and 200 $nm$) by spincoating directly on the QCM-D sensor plate. The film thicknesses were determined based on the film mass and density. The  film thickness was controlled by changing the spin speed during spincoating a xylene solution of LDPE (6 wt.\%). During spin-coating, the QCM-D plate and LDPE solution were heated with an IR lamp to prevent precipitation. Polymer coated plates were then thermally annealed under vacuum at the same conditions as bulk films before aging experiments and QCM-D measurements were performed. After UV aging, the LDPE coated QCM-D plates were rinsed with deionized water at room temperature to dissolve the polymer fragments. The water on the sample was carefully wiped off and samples were subsequently dried with nitrogen flow (50 $mL/s$). Any residual moisture was removed under vacuum at room temperature ($\SI{25} {\celsius}$) for 12 $h$. The resonance frequency of the samples was then directly measured and  converted into mass using Sauerbrey equation \citep{sauerbrey1959verwendung}. 

 
\subsection{Mechanical testing}

Specimens of as-received and aged LDPE films were cut out into dogbones and tensile tests were conducted to determine their stress-strain response before and after aging (following ASTM-D-638 standard). To subject the specimens to quasi-static loading, samples were stretched in tensile mode up to rupture at a constant strain rate of 0.004 $s^{-1}$. At least three sample tests were performed for a given exposure time to minimize uncertainty in the observed behavior.

\section{Results and Discussion} \label{results}


\subsection{Interpretation of the experimental test results} \label{exp_results_interpretation}

\subsubsection{DSC} \label{DSC discussion}

\begin{figure*}[h!bt]
    \centering
    \includegraphics[width=0.5\textwidth]{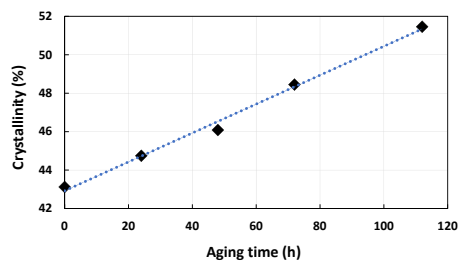}
    \caption {Evolution of crystallinity as a function of photo-oxidation aging time obtained from the DSC test.}
    \label{fig: crystal vs aging time}
\end{figure*}

Figure~\ref{fig: crystal vs aging time} presents the evolution of the crystallinity during photo-oxidation obtained based on DSC. Under the applied aging scenarios, the degree to which the crystalline part of the material gains further chain-ordering increased linearly with aging time. In fact, after just 48 $h$ of photo-oxidation, the crystallinity increased from an initial value of approximately 43\% to 46\%, totaling nearly a 7\% difference. At the end of 112 $h$, the crystallinity reached a value of nearly 52\%, which corresponds to a percent difference of about 19\% from the initial value. The shift in the crystallinity in this work is similar to the work of \cite{rodriguez2020effect} while the percentage difference is higher. The dissimilarity in the percent difference is because the initial crystalline in the work of \cite{rodriguez2020effect} was higher than LDPE here (\textit{i.e.}, 55\% compared to 43\%). This difference of the initial crystallinity can be attributed to the different annealing procedure. Particularly, the material which \cite{rodriguez2020effect} used was initially as crystallinity as our material was after 112 $h$ of UV aging. The difference in initial crystallinity could explain some of the discrepancies in material response behavior observed in our studies.  However, it is also worth mentioning that in the work of \cite{rodriguez2020effect}, the aging experiments were performed with a radiance of 1.55 $W/m^2$ compared to 125 $kW/m^2$ in our work. This means that at the end of 250 $h$ of UV aging, their samples were subjected to a total of 1.4 $MJ/m^2$ of UV radiation compared to 50 $GJ/m^2$ in our work for a duration of 112 $h$. Clearly, the extent of crystallinity change is heavily dependent on the initial composition of the material as well as exposure intensity. 

\begin{figure*}[h!bt]
    \centering
    \includegraphics[width=0.49\textwidth]{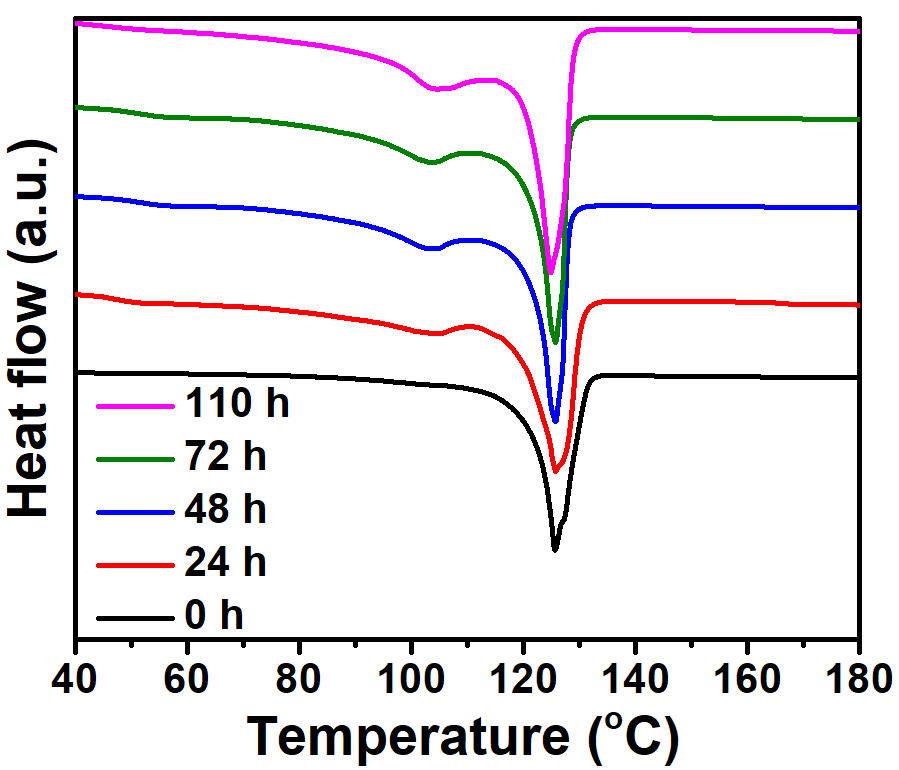}
    \caption {DSC thermograms of LDPE after varying photo-oxidation aging times.}
    \label{fig:heatflow}
\end{figure*}

The increase in crystallinity in this study contributes to a further stiffening in the material upon photo-oxidation. Whether the newly created crystallites are primary or secondary however, cannot be determined simply using Figure~\ref{fig: crystal vs aging time}. To this end, Figure~\ref{fig:heatflow} illustrates the heating thermograms of LDPE for varying photo-oxidation aging times. It can be seen that additional endothermic shoulders appeared below the melting temperature (\textit{i.e.}, $ \sim \SI{105}<\SI{125} {\celsius}$ ). However, this temperature is higher than the exposure temperature ($\SI{25} {\celsius}$). Therefore, these findings may indicate that the newly formed crystallites are secondary. However, as explained earlier in the manuscript, identifying the nature of these crystallites is not as important as recognizing that the crystallinity is inevitably expanded, and as a result, stiffness and ultimately embrittlement are significantly amplified.

While it is clear that the extent of crystallization increases linearly with aging time, it is expected, however, that the increase in crystallinity would reach a steady state some time later on in the aging process, in which case the stiffness would also reach a saturation state \citep{bhateja1983radiation}. Therefore, to account for this apparent linear dependency between the change in crystallinity and its effect on stiffness, a linear proportionality seems to be the right fit. Any changes in the crystallinity and its influence on stiffness (\textit{i.e.}, increase, constancy, or even decrease) would be appropriately captured by a linear proportionality between the initial elastic modulus and the evolution in crystallinity.

\subsubsection{QCM-D} \label{QCMD discussion}

\begin{figure*}[h!bt]
    \centering
        \includegraphics[width=0.6\linewidth]{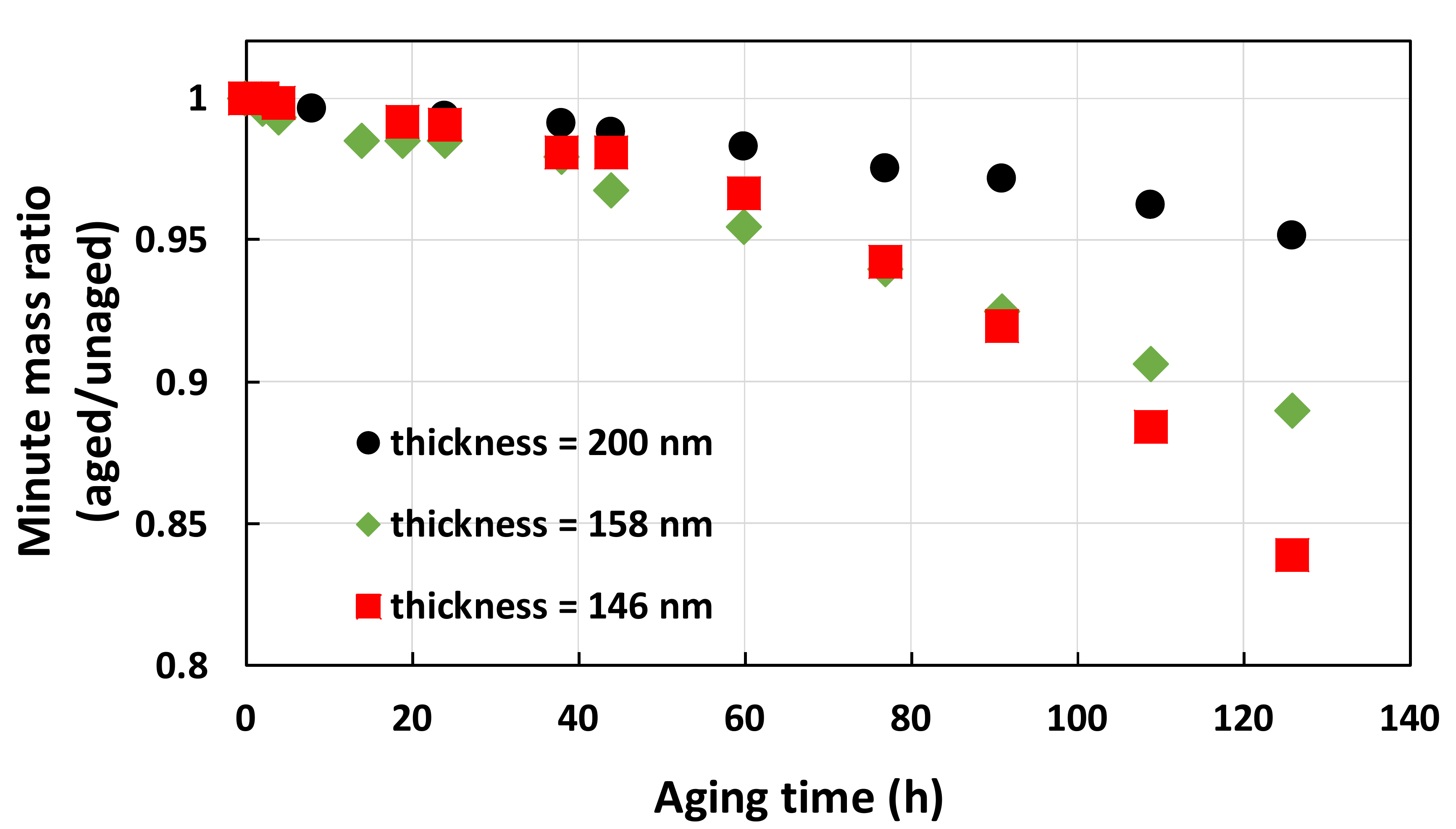}
    \caption {Evolution of the minute mass ratio between the aged and unaged samples as a function of photo-oxidation aging time obtained from the QCM-D test. The minute mass ratio is presented for three film thicknesses; 200~$nm$, 158~$nm$, and 146~$nm$ represented by circles, hexagons, and squares, respectively.} 
    \label{fig: massloss}
\end{figure*}

Figure~\ref{fig: massloss} illustrates the evolution of the minute mass ratio with respect to aging time measured by the QCM-D for three different LDPE film thicknesses. The 200-$nm$-thick film experienced a nearly 5\% weight loss after 120 $h$ of UV aging. On the other hand, the two remaining thinner films experienced more weight loss under the same aging duration (\textit{i.e.}, up to 15\% for the 146-$nm$-thick film). Nevertheless, given that the LDPE coated QCM-D plates were rinsed with deinonized water, a 5\% mass loss at 112 $h$ of UV aging with a dose rate equal to 125 $kW/m^2$ is remarkable. 
Indeed, microplastics are found in exuberant amounts largely due to plastic-fragmentation caused by the exposure of plastics to environmental perturbations such as UV radiation.

\subsubsection{Mechanical testing}

\begin{figure*}[h!bt]
    \centering
        \includegraphics[width=0.7\linewidth]{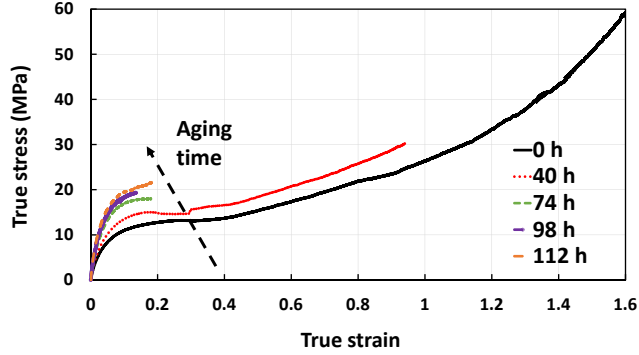}
    \caption {Average engineering stress-strain curves from each group of at least three replicates corresponding to different photo-oxidation aging times (\textit{i.e.}, 0, 40, 74, 98, and 112 $h$).} 
    \label{fig: avgEngSS}
\end{figure*}

Tensile stress-strain curves for LDPE were obtained for the aging times considered in this work (\textit{i.e.}, 0, 48, 74, 98, and 112 $h$). At least three replicates were tested for each aging time. The averages amongst each group of replicates were taken and the result were plotted in Figure~\ref{fig: avgEngSS}. It can be seen that both the initial stiffness and the yield stress increased by increasing aging time. On the other hand, the films showed a substantial reduction in ductility. The increase in the initial stiffness and yield stress are indicative of chemi-crystallization and chain crosslinking. The loss of ductility is indicative of a reduction in the molecular weight. The observed effects of photo-oxidation on the mechanical performance of LDPE (\textit{i.e.}, increase in initial stiffness and yield stress and decrease in ductility) are expected and supported by the characterizations of DSC and QCM-D. On the one hand, the expansion of the crystalline domain at the expense of its amorphous counterpart after long aging times explains the enhanced initial stiffness and yield stress. On the other hand, the minute mass loss determined by the QCM-D indicates LDPE degradation during photo-oxidation which reduces chain integrity and compromises the mechanical response, causing a substantial decrease in material ductility over exposure time. Here, it is worth mentioning that although the weight loss in bulk polymer films may not be comparable with that of thin films under the same aging conditions, 
the loss of LDPE chain integrity after photo-oxidation is expected be comparable for both thin and bulk polymer films due to the good UV light transmittance in polyethylene at thicknesses lower than 80 $\mu m$. 
Therefore, other than the crystallinity change determined by DSC, the reduced ductility after photo-oxidation is also explained by the mass loss monitored by QCM-D .


\subsection{Prediction capability of the proposed constitutive framework}

In this section, prediction capability of the proposed constitutive framework is discussed. 
The constitutive framework was numerically implemented into a Matlab code for the case of uniaxial tensile loading. The true stress-strain measures were used to describe the deformation behavior of the material. 

\begin{figure*}[h!bt]
    \centering
        \includegraphics[width=0.7\linewidth]{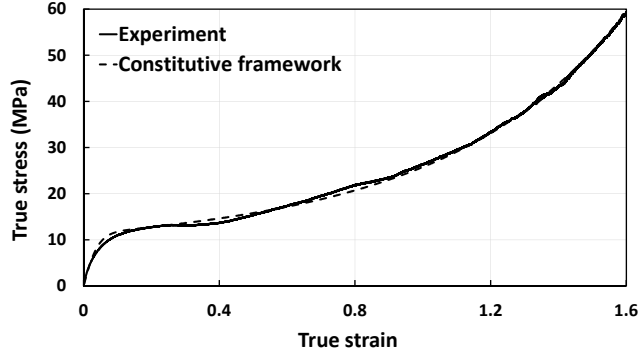}
    \caption {True stress-strain curve of unaged LDPE representing the experiment versus numerical response based on the constitutive framework.} 
    \label{fig: unagedprediction}
\end{figure*}


To begin, the unaged tensile test was used to determine the material properties of the unaged film involved in the constitutive framework (\textit{i.e.}, $E_0$, $s_0$, $\Delta G_{a0}$, $\mu_0$, and $N_{k0}$). 
Specifically, the elastic modulus $E_0$ and the rubber modulus $\mu_0$ were determined as the slope of the low-strain and the large-strain regions of the tensile stress-strain curve, respectively. Additionally, the athermal shear strength $s_0$, the activation energy $\Delta G_{a0}$, and the number of Kuhn monomers $N_{k0}$ were determined by fitting the unaged stress-strain curve to the numerical response. An alternative method to obtain $s_0$ and $\Delta G_{a0}$ is through conducting tensile tests at varying strain rates and using equation~\ref{eq: viscoplastic shear strain} to back-calculate the values of the two properties. 
Table~\ref{Model parameters} summarizes the obtained unaged material properties and Figure~\ref{fig: unagedprediction} illustrates the comparison between the unaged numerical and experimental true stress-strain responses. 
Once the unaged material properties were determined, their evolution according to the proposed evolution functions (\textit{i.e.}, Equations~\ref{eq: evolution of E}, ~\ref{eq: evolution of s}, and ~\ref{eq: evolution of delG}) for any aging time could be readily acquired.

\begin{table}[h!bt]
\centering
\small
\caption{Material properties for the unaged LDPE specimen.}
\label{Model parameters}
    \begin{tabular}{cccc}
    \hline
        Branch contribution & Parameters & Equation & Values \\ \hline
        \multirow{5}{*}{\textbf{Intermolecular resistance}} & Elastic modulus ($MPa$), $E_0$ & \ref{eq: fourth-order elastic stiffness tensor} & 596 \\
        {} & Poisson's ratio, $\nu$ & \ref{eq: fourth-order elastic stiffness tensor} & 0.49 \\
        {} & Pre-exponential factor ($s^{-1}$), $\dot\gamma_0$  & \ref{eq: viscoplastic shear strain} & $1.75\times 10^6$  \\
        {} & Athermal shear strength ($MPa$), $s_0$ & \ref{eq: viscoplastic shear strain} & 155 \\
        {} & Activation energy ($J$), $\Delta G_{a0}$ & \ref{eq: viscoplastic shear strain} & $8.5\times 10^{-17}$ \\
        \hline
        \multirow{3}{*}{\textbf{Network resistance}} & Rubbery modulus ($MPa$), $\mu_0$ & \ref{eq: Arruda-Boyce stress tensor} & 2.3 \\
        {} & Number of Kuhn monomers, $N_{k0}$ & \ref{eq: Arruda-Boyce stress tensor} & 100 \\
        {} & Relaxation parameter ($MPa^{-1}$), C & \ref{eq: flow shear strain rate} & $8 \times 10^{-8}$ \\ \hline
    \end{tabular}
\end{table}

\begin{figure*}[hbt!p]
    \centering
    \subfloat[All aging times]{%
        \includegraphics[width=0.8\textwidth]{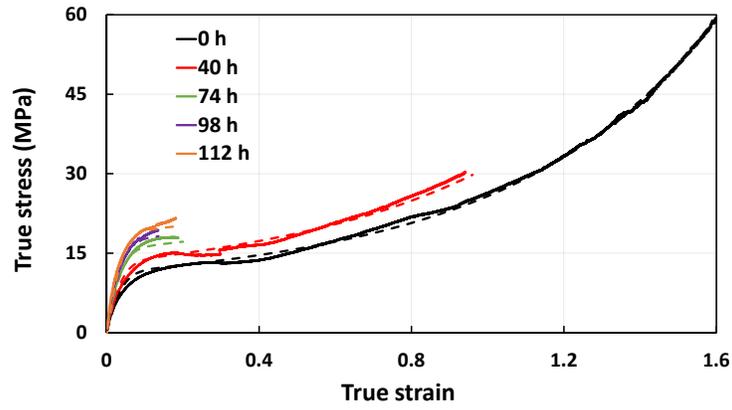}
        \label{fig:all}
        } \\
    \hfill%
    \subfloat[Aged 40 h]{%
        \includegraphics[width=0.49\textwidth]{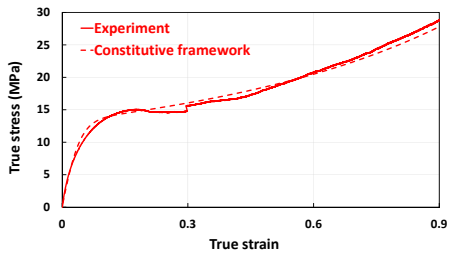}
        \label{fig:40}
        }%
    \hfill%
    \subfloat[Aged 74 h]{%
        \includegraphics[width=0.49\textwidth]{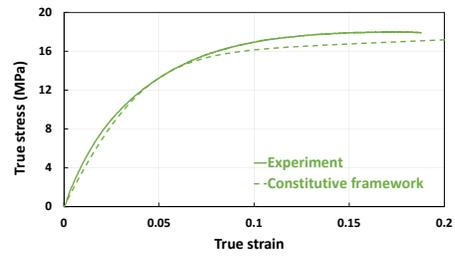}
        \label{fig:74}
        }%
    \vspace*{-1cm}
    \hfill%
    \subfloat[Aged 98 h]{%
        \includegraphics[width=0.49\textwidth]{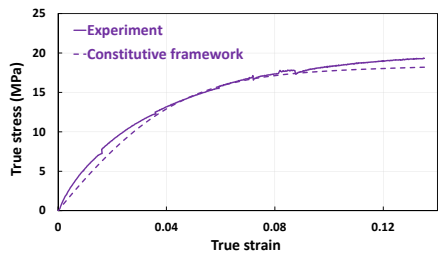}
        \label{fig:98}
        }%
    \hfill%
    \subfloat[Aged 112 h]{%
        \includegraphics[width=0.49\textwidth]{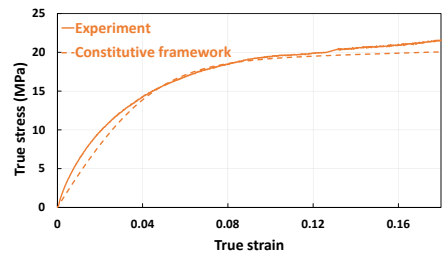}
        \label{fig:112}
        }%
    \caption {Constitutive framework prediction versus experimental results for the varying aging times considered in this work.}
    \label{fig: agedprediction}
\end{figure*}

Figure~\ref{fig: agedprediction} demonstrates the comparison between the experimental tensile test and the developed constitutive framework for photo-oxidatively aged LDPE under varying exposure times. Particularly, Figure~\ref{fig:all} summarizes the comparison results from all of the considered aging times, while Figures~\ref{fig:40}-\ref{fig:112} focus on the response predictions at each aging time separately to better appreciate the accuracy of predictions. A very good prediction could be obtained for all aging times. Both the initial modulus and the yield stress accurately matched with the experimental results for the varying aging times considered in this study. 

\subsection{Discussion}
Physio-chemically-motivated evolution functions based on LDPE film minute mass loss and crystallinity changes are important for accurate prediction of photo-oxidation effects on the mechanical response of LDPE. Obtaining the minute mass loss evolution functions is challenging primarily due to the minor weight loss and negligible mass change for thick polymer films. Therefore, to accurately measure the mass loss during aging, the use of thin films is an inevitable selection. However, when it comes to thin films, the thickness effect on the film behavior is considerable. For instance, the film thickness has an observable effect on the glass transition temperature of polymer when it is around 100~$nm$ due to the increased surface effect of thin films \citep{peter2006thickness}. Therefore, to consider the minor mass loss for thick films and avoid spurious thickness effects for thin films, LDPE films with thickness less than 1 $\mu m$ but greater than 100~$nm$ can be used to amplify the mass loss under the aging conditions and accurately measure the change of mass.  
Indeed, an increase in the film thickness in the QCM-D measurement may further improve the accuracy and reliability of result predictions; nonetheless, based on the mass loss evolution of films with thickness equal to 200~$nm$, the proposed constitutive framework already can predict the mechanical responses of photo-oxidatively aged semi-crystalline LDPE very well as it can be seen in Figure~\ref{fig: agedprediction}. 

In addition to the minute mass loss, crystallinity behavior also merits careful attention. In particular, the initial crystallinity determined in this work was relatively lower than the one reported by \cite{rodriguez2020effect}. This particular difference in the measured initial crystallinity may have had a significant contribution to the differences observed in the mechanical responses of unaged and aged samples between our study and the work by \cite{rodriguez2020effect}. Indeed, after 112 $h$ of UV radiation with an intensity of 125 $kW/m^2$, our material was just as crystalline as the material used in \cite{rodriguez2020effect} initially was (\textit{i.e.}, before any UV exposure). Therefore, comparison between both of these works should be approached with care to make meaningful conclusions regarding crystallinity change effects on the evolution of LDPE material properties.

The ability of the developed constitutive framework to accurately predict the mechanical test results of aged LDPE independently of any mechanical tests constitutes the important contribution of the proposed framework. Indeed, many, if not all of the existing works use several mechanical tests to fit and obtain fitting parameters that carry no physical meaning within the overall material behavior \citep{ayoub2020modeling,lamnii2021experimental,belbachir2010modelling}. Doing so renders the constitutive approach essentially a fitting algorithm with numerous fitting parameters that applies only to the specific problem for which calibration was performed. In contrast, developing a general framework which is physics- and chemistry-based and is comprehensive in its prediction capability is more reliable. As demonstrated throughout the manuscript, our constitutive framework predicts the responses of aged LDPE without conducting any further fitting to the mechanical test results on aged samples. More so, it can predict the responses of photo-oxidatively aged semi-crystalline polymers with high accuracy. Therefore, the developed physio-chemically-motivated framework is unique and unprecedented.

\section{Concluding remarks} \label{conclusion}

We developed a purely physio-chemically-based constitutive framework to predict the mechanical performance of semi-crystalline LDPE in response to photo-oxidative aging. In contrast to all modeling efforts in the literature, we based the evolution of the macromechanical properties in response to photo-oxidation on the chemically verified processes responsible for material degradation. In doing so, we eliminated the need to employ extra fitting parameters which carry no physical meaning. The framework was based on modifying the constitutive equations of \cite{boyce2000constitutive} to incorporate the effects of crystallinity evolution and minute mass ratio change in modifying the elasto-viscoplastic material properties. The crystallinity change was measured with DSC whereas the minute mass loss was measured with QCM-D. The use of QCM-D as a characterization technique for photo-oxidation investigation was validated through comparison between numerical and experimental tensile test results. Particularly, we showed that the minute mass ratio can be directly related to polymer stiffening and increase in yield stress and conjectured appropriate evolution functions for the material properties probing polymer response to chemical changes. These chemical characterizations (\textit{i.e.}, DSC and QCM-D) determined the changes in the physio-chemical structure of the material and bridged the gap between molecular network evolution and its effect on the overall macroscopic mechanical changes. The developed constitutive framework could predict the mechanical responses of photo-oxidatively aged LDPE independently of mechanical tests on aged specimens with high accuracy. It thus provides a one-to-one mapping between chemistry-based quantities (\textit{i.e.}, crystallinity and minute mass ratio) and physics-based macroscopic variables (\textit{i.e.}, elasto-viscoplastic mechanical properties of the material).

A possible future investigation is to implement the developed three-dimensional constitutive framework into a finite element software that allows for various additional considerations (\textit{e.g.}, more complex load states, coupled chemo-mechanical diffusion problem, etc). This can be realized through the incorporation of kinetics equations based on the chemical characterizations presented in this work (\textit{i.e.}, crystallinity change and evolution of mass loss) coupled with a diffusion-deformation problem. Another possible future study is to incorporate damage into the developed constitutive framework to capture photo-oxidation-induced failure of aged polymers. 
Consideration of such important developments is particularly essential in ensuring durable polymer design and active environment protection and is the topic of future work by the authors.

\section*{Acknowledgement}
The authors gratefully acknowledge the support from the National Science Foundation under the award number CMMI-1914565.

\section*{Appendix. A} \label{appendix}
The determinant of the total deformation gradient $\mathbf{F}$, det$\mathbf{F}$, can be multiplicatively decomposed into elastic and plastic components as det $\mathbf{F}=J^e J^p > 1$, in which we define $J^e = $ det$\mathbf{F}^e$ and $J^p = $ det$\mathbf{F}^p$. Assuming that plastic flow is volume preserving (\textit{i.e.}, incompressible), we write $J^p =$ det$\mathbf{F}^p = 1$. Note that the decomposition used in Equations~\ref{lee decomp} indicates that there exists an intermediate configuration (\textit{i.e.}, a relaxed configuration) between the undeformed and the current configurations. The relaxed configuration is assumed to be obtained from the current configuration by unloading through the inverse of the elastic part of the deformation gradients. 

Additionally, we can use the polar decomposition of the deformation gradients Equations~\ref{lee decomp} and write \citep{gurtin2005decomposition}:
\begin{align} \label{polar decomp I}
    \mathbf{F}_I = \mathbf{V}_I^e \mathbf{R}_I^e \mathbf{V}_I^p \mathbf{R}_I^p
\end{align}
\begin{align} \label{polar decomp N}
    \mathbf{F}_N = \mathbf{V}_N^e \mathbf{R}_N^e \mathbf{V}_N^p \mathbf{R}_N^p
\end{align}
where $\mathbf{V}$ and $\mathbf{R}$ refer to the stretch (symmetric) and rotation (orthogonal) parts of the corresponding deformation gradient, respectively.

The velocity gradients $\mathbf{L}_I = \dot{\mathbf{F}}_I {\mathbf{F}}_I^{-1}$ for branch I and $\mathbf{L}_N = \dot{\mathbf{F}}_N {\mathbf{F}}_N^{-1}$ for branch N can be computed as follows: 
\begin{align} \label{vel gradient I}
    \mathbf{L}_I = \dot{\mathbf{F}}_I {\mathbf{F}}_I^{-1} = \dot{\mathbf{F}}_I^e {\mathbf{F}_I^e}^{-1} + \mathbf{F}_I^e \dot{\mathbf{F}}_I^p {\mathbf{F}_I^p}^{-1}{\mathbf{F}_I^e}^{-1} = \mathbf{L}_I^e + \mathbf{L}_I^p
\end{align}
\begin{align} \label{vel gradient N}
    \mathbf{L}_N = \dot{\mathbf{F}}_N {\mathbf{F}}_N^{-1} = \dot{\mathbf{F}}_N^e {\mathbf{F}_N^e}^{-1} + \mathbf{F}_N^e \dot{\mathbf{F}}_N^p {\mathbf{F}_N^p}^{-1}{\mathbf{F}_N^e}^{-1} = \mathbf{L}_N^e + \mathbf{L}_N^p
\end{align}
The plastic components of the velocity gradients $\mathbf{L}_I^p = \mathbf{F}_I^e \dot{\mathbf{F}}_I^p {\mathbf{F}_I^p}^{-1}{\mathbf{F}_I^e}^{-1} $ and $\mathbf{L}_N^p = \mathbf{F}_N^e \dot{\mathbf{F}}_N^p {\mathbf{F}_N^p}^{-1}{\mathbf{F}_N^e}^{-1}$ can also further be decomposed into their symmetric and skew parts as follows: 
\begin{align} \label{sym I}
    \mathbf{L}_I^p = \mathbf{D}_I^p + \mathbf{W}_I^p 
\end{align}
\begin{align} \label{sym N}
    \mathbf{L}_N^p = \mathbf{D}_N^p + \mathbf{W}_N^p 
\end{align}
where $\mathbf{D}_I^p$ and $\mathbf{D}_N^p$ are the rates of inelastic deformation, and $\mathbf{W}_I^p$ and $\mathbf{W}_N^p$ are the inelastic spins which are assumed, without loss of generality, to be equal to zero (\textit{i.e.}, irrotational).

\bibliography{References}

\begin{thebibliography}{59}
\expandafter\ifx\csname natexlab\endcsname\relax\def\natexlab#1{#1}\fi
\providecommand{\url}[1]{\texttt{#1}}
\providecommand{\href}[2]{#2}
\providecommand{\path}[1]{#1}
\providecommand{\DOIprefix}{doi:}
\providecommand{\ArXivprefix}{arXiv:}
\providecommand{\URLprefix}{URL: }
\providecommand{\Pubmedprefix}{pmid:}
\providecommand{\doi}[1]{\href{http://dx.doi.org/#1}{\path{#1}}}
\providecommand{\Pubmed}[1]{\href{pmid:#1}{\path{#1}}}
\providecommand{\bibinfo}[2]{#2}
\ifx\xfnm\relax \def\xfnm[#1]{\unskip,\space#1}\fi
\bibitem[{Abdelaziz et~al.(2019)Abdelaziz, Ayoub, Colin, Benhassine \&
  Mouwakeh}]{abdelaziz2019new}
\bibinfo{author}{Abdelaziz, M.~N.}, \bibinfo{author}{Ayoub, G.},
  \bibinfo{author}{Colin, X.}, \bibinfo{author}{Benhassine, M.}, \&
  \bibinfo{author}{Mouwakeh, M.} (\bibinfo{year}{2019}).
\newblock \bibinfo{title}{New developments in fracture of rubbers: Predictive
  tools and influence of thermal aging}.
\newblock {\it \bibinfo{journal}{International Journal of Solids and
  Structures}\/},  {\it \bibinfo{volume}{165}\/}, \bibinfo{pages}{127--136}.
\bibitem[{Abdul-Hameed et~al.(2014)Abdul-Hameed, Messager, Za{\"\i}ri \&
  Na{\"\i}t-Abdelaziz}]{abdul2014large}
\bibinfo{author}{Abdul-Hameed, H.}, \bibinfo{author}{Messager, T.},
  \bibinfo{author}{Za{\"\i}ri, F.}, \& \bibinfo{author}{Na{\"\i}t-Abdelaziz,
  M.} (\bibinfo{year}{2014}).
\newblock \bibinfo{title}{Large-strain viscoelastic--viscoplastic constitutive
  modeling of semi-crystalline polymers and model identification by
  deterministic/evolutionary approach}.
\newblock {\it \bibinfo{journal}{Computational Materials Science}\/},  {\it
  \bibinfo{volume}{90}\/}, \bibinfo{pages}{241--252}.
\bibitem[{Ahzi et~al.(2003)Ahzi, Makradi, Gregory \& Edie}]{ahzi2003modeling}
\bibinfo{author}{Ahzi, S.}, \bibinfo{author}{Makradi, A.},
  \bibinfo{author}{Gregory, R.}, \& \bibinfo{author}{Edie, D.}
  (\bibinfo{year}{2003}).
\newblock \bibinfo{title}{Modeling of deformation behavior and strain-induced
  crystallization in poly (ethylene terephthalate) above the glass transition
  temperature}.
\newblock {\it \bibinfo{journal}{Mechanics of Materials}\/},  {\it
  \bibinfo{volume}{35}\/}, \bibinfo{pages}{1139--1148}.
\bibitem[{Anand \& Gurtin(2003)}]{anand2003theory}
\bibinfo{author}{Anand, L.}, \& \bibinfo{author}{Gurtin, M.~E.}
  (\bibinfo{year}{2003}).
\newblock \bibinfo{title}{A theory of amorphous solids undergoing large
  deformations, with application to polymeric glasses}.
\newblock {\it \bibinfo{journal}{International Journal of Solids and
  Structures}\/},  {\it \bibinfo{volume}{40}\/}, \bibinfo{pages}{1465--1487}.
\bibitem[{Arruda \& Boyce(1993)}]{arruda1993three}
\bibinfo{author}{Arruda, E.~M.}, \& \bibinfo{author}{Boyce, M.~C.}
  (\bibinfo{year}{1993}).
\newblock \bibinfo{title}{A three-dimensional constitutive model for the large
  stretch behavior of rubber elastic materials}.
\newblock {\it \bibinfo{journal}{Journal of the Mechanics and Physics of
  Solids}\/},  {\it \bibinfo{volume}{41}\/}, \bibinfo{pages}{389--412}.
\bibitem[{Arruda et~al.(1995)Arruda, Boyce \& Jayachandran}]{arruda1995effects}
\bibinfo{author}{Arruda, E.~M.}, \bibinfo{author}{Boyce, M.~C.}, \&
  \bibinfo{author}{Jayachandran, R.} (\bibinfo{year}{1995}).
\newblock \bibinfo{title}{Effects of strain rate, temperature and
  thermomechanical coupling on the finite strain deformation of glassy
  polymers}.
\newblock {\it \bibinfo{journal}{Mechanics of Materials}\/},  {\it
  \bibinfo{volume}{19}\/}, \bibinfo{pages}{193--212}.
\bibitem[{Ayoub et~al.(2020)Ayoub, Rodriguez, Mansoor \&
  Colin}]{ayoub2020modeling}
\bibinfo{author}{Ayoub, G.}, \bibinfo{author}{Rodriguez, A.},
  \bibinfo{author}{Mansoor, B.}, \& \bibinfo{author}{Colin, X.}
  (\bibinfo{year}{2020}).
\newblock \bibinfo{title}{Modeling the visco-hyperelastic--viscoplastic
  behavior of photodegraded semi-crystalline low-density polyethylene films}.
\newblock {\it \bibinfo{journal}{International Journal of Solids and
  Structures}\/},  {\it \bibinfo{volume}{204}\/}, \bibinfo{pages}{187--198}.
\bibitem[{Ayoub et~al.(2010)Ayoub, Za{\"\i}ri, Na{\"\i}t-Abdelaziz \&
  Gloaguen}]{ayoub2010modelling}
\bibinfo{author}{Ayoub, G.}, \bibinfo{author}{Za{\"\i}ri, F.},
  \bibinfo{author}{Na{\"\i}t-Abdelaziz, M.}, \& \bibinfo{author}{Gloaguen,
  J.~M.} (\bibinfo{year}{2010}).
\newblock \bibinfo{title}{Modelling large deformation behaviour under
  loading--unloading of semicrystalline polymers: application to a high density
  polyethylene}.
\newblock {\it \bibinfo{journal}{International Journal of Plasticity}\/},  {\it
  \bibinfo{volume}{26}\/}, \bibinfo{pages}{329--347}.
\bibitem[{Bardenhagen et~al.(1997)Bardenhagen, Stout \&
  Gray}]{bardenhagen1997three}
\bibinfo{author}{Bardenhagen, S.~G.}, \bibinfo{author}{Stout, M.~G.}, \&
  \bibinfo{author}{Gray, G.~T.} (\bibinfo{year}{1997}).
\newblock \bibinfo{title}{Three-dimensional, finite deformation, viscoplastic
  constitutive models for polymeric materials}.
\newblock {\it \bibinfo{journal}{Mechanics of Materials}\/},  {\it
  \bibinfo{volume}{25}\/}, \bibinfo{pages}{235--253}.
\bibitem[{Belbachir et~al.(2010)Belbachir, Za{\"\i}ri, Ayoub, Maschke,
  Na{\"\i}t-Abdelaziz, Gloaguen, Benguediab \&
  Lefebvre}]{belbachir2010modelling}
\bibinfo{author}{Belbachir, S.}, \bibinfo{author}{Za{\"\i}ri, F.},
  \bibinfo{author}{Ayoub, G.}, \bibinfo{author}{Maschke, U.},
  \bibinfo{author}{Na{\"\i}t-Abdelaziz, M.}, \bibinfo{author}{Gloaguen, J.~M.},
  \bibinfo{author}{Benguediab, M.}, \& \bibinfo{author}{Lefebvre, J.~M.}
  (\bibinfo{year}{2010}).
\newblock \bibinfo{title}{Modelling of photodegradation effect on
  elastic--viscoplastic behaviour of amorphous polylactic acid films}.
\newblock {\it \bibinfo{journal}{Journal of the Mechanics and Physics of
  Solids}\/},  {\it \bibinfo{volume}{58}\/}, \bibinfo{pages}{241--255}.
\bibitem[{Bergmann et~al.(2019)Bergmann, M{\"u}tzel, Primpke, Tekman, Trachsel
  \& Gerdts}]{bergmann2019white}
\bibinfo{author}{Bergmann, M.}, \bibinfo{author}{M{\"u}tzel, S.},
  \bibinfo{author}{Primpke, S.}, \bibinfo{author}{Tekman, M.~B.},
  \bibinfo{author}{Trachsel, J.}, \& \bibinfo{author}{Gerdts, G.}
  (\bibinfo{year}{2019}).
\newblock \bibinfo{title}{White and wonderful? microplastics prevail in snow
  from the alps to the arctic}.
\newblock {\it \bibinfo{journal}{Science Advances}\/},  {\it
  \bibinfo{volume}{5}\/}, \bibinfo{pages}{eaax1157}.
\bibitem[{Bhateja(1983)}]{bhateja1983radiation}
\bibinfo{author}{Bhateja, S.} (\bibinfo{year}{1983}).
\newblock \bibinfo{title}{Radiation-induced crystallinity changes in linear
  polyethylene: Influence of aging}.
\newblock {\it \bibinfo{journal}{Journal of Applied Polymer Science}\/},  {\it
  \bibinfo{volume}{28}\/}, \bibinfo{pages}{861--872}.
\bibitem[{Boyce et~al.(2000)Boyce, Socrate \& Llana}]{boyce2000constitutive}
\bibinfo{author}{Boyce, M.}, \bibinfo{author}{Socrate, S.}, \&
  \bibinfo{author}{Llana, P.} (\bibinfo{year}{2000}).
\newblock \bibinfo{title}{Constitutive model for the finite deformation
  stress--strain behavior of poly (ethylene terephthalate) above the glass
  transition}.
\newblock {\it \bibinfo{journal}{Polymer}\/},  {\it \bibinfo{volume}{41}\/},
  \bibinfo{pages}{2183--2201}.
\bibitem[{Boyce et~al.(1988)Boyce, Parks \& Argon}]{boyce1988large}
\bibinfo{author}{Boyce, M.~C.}, \bibinfo{author}{Parks, D.~M.}, \&
  \bibinfo{author}{Argon, A.~S.} (\bibinfo{year}{1988}).
\newblock \bibinfo{title}{Large inelastic deformation of glassy polymers. part
  i: rate dependent constitutive model}.
\newblock {\it \bibinfo{journal}{Mechanics of Materials}\/},  {\it
  \bibinfo{volume}{7}\/}, \bibinfo{pages}{15--33}.
\bibitem[{Brandon et~al.(2016)Brandon, Goldstein \& Ohman}]{brandon2016long}
\bibinfo{author}{Brandon, J.}, \bibinfo{author}{Goldstein, M.}, \&
  \bibinfo{author}{Ohman, M.~D.} (\bibinfo{year}{2016}).
\newblock \bibinfo{title}{Long-term aging and degradation of microplastic
  particles: comparing in situ oceanic and experimental weathering patterns}.
\newblock {\it \bibinfo{journal}{Marine Pollution Bulletin}\/},  {\it
  \bibinfo{volume}{110}\/}, \bibinfo{pages}{299--308}.
\bibitem[{Breche et~al.(2016{\natexlab{a}})Breche, Chagnon, Machado, Girard,
  Nottelet, Garric \& Favier}]{breche2016mechanical}
\bibinfo{author}{Breche, Q.}, \bibinfo{author}{Chagnon, G.},
  \bibinfo{author}{Machado, G.}, \bibinfo{author}{Girard, E.},
  \bibinfo{author}{Nottelet, B.}, \bibinfo{author}{Garric, X.}, \&
  \bibinfo{author}{Favier, D.} (\bibinfo{year}{2016}{\natexlab{a}}).
\newblock \bibinfo{title}{Mechanical behaviour's evolution of a pla-b-peg-b-pla
  triblock copolymer during hydrolytic degradation}.
\newblock {\it \bibinfo{journal}{Journal of the Mechanical Behavior of
  Biomedical Materials}\/},  {\it \bibinfo{volume}{60}\/},
  \bibinfo{pages}{288--300}.
\bibitem[{Breche et~al.(2016{\natexlab{b}})Breche, Chagnon, Machado, Nottelet,
  Garric, Girard \& Favier}]{breche2016non}
\bibinfo{author}{Breche, Q.}, \bibinfo{author}{Chagnon, G.},
  \bibinfo{author}{Machado, G.}, \bibinfo{author}{Nottelet, B.},
  \bibinfo{author}{Garric, X.}, \bibinfo{author}{Girard, E.}, \&
  \bibinfo{author}{Favier, D.} (\bibinfo{year}{2016}{\natexlab{b}}).
\newblock \bibinfo{title}{A non-linear viscoelastic model to describe the
  mechanical behavior's evolution of biodegradable polymers during hydrolytic
  degradation}.
\newblock {\it \bibinfo{journal}{Polymer Degradation and Dtability}\/},  {\it
  \bibinfo{volume}{131}\/}, \bibinfo{pages}{145--156}.
\bibitem[{Carrasco et~al.(2001)Carrasco, Pagès, Pascual \&
  Colom}]{CARRASCO20011457}
\bibinfo{author}{Carrasco, F.}, \bibinfo{author}{Pagès, P.},
  \bibinfo{author}{Pascual, S.}, \& \bibinfo{author}{Colom, X.}
  (\bibinfo{year}{2001}).
\newblock \bibinfo{title}{Artificial aging of high-density polyethylene by
  ultraviolet irradiation}.
\newblock {\it \bibinfo{journal}{European Polymer Journal}\/},  {\it
  \bibinfo{volume}{37}\/}, \bibinfo{pages}{1457 -- 1464}.
\bibitem[{Celina(2013)}]{celina2013review}
\bibinfo{author}{Celina, M.~C.} (\bibinfo{year}{2013}).
\newblock \bibinfo{title}{Review of polymer oxidation and its relationship with
  materials performance and lifetime prediction}.
\newblock {\it \bibinfo{journal}{Polymer Degradation and Stability}\/},  {\it
  \bibinfo{volume}{98}\/}, \bibinfo{pages}{2419--2429}.
\bibitem[{Cohen(1991)}]{cohen1991pade}
\bibinfo{author}{Cohen, A.} (\bibinfo{year}{1991}).
\newblock \bibinfo{title}{A pad{\'e} approximant to the inverse langevin
  function}.
\newblock {\it \bibinfo{journal}{Rheologica acta}\/},  {\it
  \bibinfo{volume}{30}\/}, \bibinfo{pages}{270--273}.
\bibitem[{Cundiff et~al.(2020)Cundiff, Madi \& Benzerga}]{cundiff2020photo}
\bibinfo{author}{Cundiff, K.}, \bibinfo{author}{Madi, Y.}, \&
  \bibinfo{author}{Benzerga, A.} (\bibinfo{year}{2020}).
\newblock \bibinfo{title}{Photo-oxidation of semicrystalline polymers: Damage
  nucleation versus growth}.
\newblock {\it \bibinfo{journal}{Polymer}\/},  {\it \bibinfo{volume}{188}\/},
  \bibinfo{pages}{122090}.
\bibitem[{Da~Costa et~al.(2018)Da~Costa, Nunes, Santos, Girao, Duarte \&
  Rocha-Santos}]{da2018degradation}
\bibinfo{author}{Da~Costa, J.~P.}, \bibinfo{author}{Nunes, A.~R.},
  \bibinfo{author}{Santos, P.~S.}, \bibinfo{author}{Girao, A.~V.},
  \bibinfo{author}{Duarte, A.~C.}, \& \bibinfo{author}{Rocha-Santos, T.}
  (\bibinfo{year}{2018}).
\newblock \bibinfo{title}{Degradation of polyethylene microplastics in
  seawater: Insights into the environmental degradation of polymers}.
\newblock {\it \bibinfo{journal}{Journal of Environmental Science and Health,
  Part A}\/},  {\it \bibinfo{volume}{53}\/}, \bibinfo{pages}{866--875}.
\bibitem[{Dupaix \& Krishnan(2006)}]{dupaix2006constitutive}
\bibinfo{author}{Dupaix, R.~B.}, \& \bibinfo{author}{Krishnan, D.}
  (\bibinfo{year}{2006}).
\newblock \bibinfo{title}{A constitutive model for strain-induced
  crystallization in poly (ethylene terephthalate)(pet) during finite strain
  load-hold simulations}.
\newblock {\it \bibinfo{journal}{Journal of Engineering Materials and
  Technology}\/},  {\it \bibinfo{volume}{128}\/}, \bibinfo{pages}{28--33}.
\bibitem[{Fayolle et~al.(2008)Fayolle, Richaud, Colin \&
  Verdu}]{fayolle2008degradation}
\bibinfo{author}{Fayolle, B.}, \bibinfo{author}{Richaud, E.},
  \bibinfo{author}{Colin, X.}, \& \bibinfo{author}{Verdu, J.}
  (\bibinfo{year}{2008}).
\newblock \bibinfo{title}{degradation-induced embrittlement in semi-crystalline
  polymers having their amorphous phase in rubbery state}.
\newblock {\it \bibinfo{journal}{Journal of Materials Science}\/},  {\it
  \bibinfo{volume}{43}\/}, \bibinfo{pages}{6999--7012}.
\bibitem[{Guo \& Wang(2019)}]{guo2019chemical}
\bibinfo{author}{Guo, X.}, \& \bibinfo{author}{Wang, J.}
  (\bibinfo{year}{2019}).
\newblock \bibinfo{title}{The chemical behaviors of microplastics in marine
  environment: A review}.
\newblock {\it \bibinfo{journal}{Marine Pollution Bulletin}\/},  {\it
  \bibinfo{volume}{142}\/}, \bibinfo{pages}{1--14}.
\bibitem[{Gurtin \& Anand(2005)}]{gurtin2005decomposition}
\bibinfo{author}{Gurtin, M.~E.}, \& \bibinfo{author}{Anand, L.}
  (\bibinfo{year}{2005}).
\newblock \bibinfo{title}{The decomposition f= fefp, material symmetry, and
  plastic irrotationality for solids that are isotropic-viscoplastic or
  amorphous}.
\newblock {\it \bibinfo{journal}{International Journal of Plasticity}\/},  {\it
  \bibinfo{volume}{21}\/}, \bibinfo{pages}{1686--1719}.
\bibitem[{Han \& Pan(2009)}]{han2009model}
\bibinfo{author}{Han, X.}, \& \bibinfo{author}{Pan, J.} (\bibinfo{year}{2009}).
\newblock \bibinfo{title}{A model for simultaneous crystallisation and
  biodegradation of biodegradable polymers}.
\newblock {\it \bibinfo{journal}{Biomaterials}\/},  {\it
  \bibinfo{volume}{30}\/}, \bibinfo{pages}{423--430}.
\bibitem[{Hedir et~al.(2020)Hedir, Moudoud, Lamrous, Rondot, Jbara \&
  Dony}]{Hedir2020}
\bibinfo{author}{Hedir, A.}, \bibinfo{author}{Moudoud, M.},
  \bibinfo{author}{Lamrous, O.}, \bibinfo{author}{Rondot, S.},
  \bibinfo{author}{Jbara, O.}, \& \bibinfo{author}{Dony, P.}
  (\bibinfo{year}{2020}).
\newblock \bibinfo{title}{Ultraviolet radiation aging impact on physicochemical
  properties of crosslinked polyethylene cable insulation}.
\newblock {\it \bibinfo{journal}{Journal of Applied Polymer Science}\/},  {\it
  \bibinfo{volume}{137}\/}, \bibinfo{pages}{48575}.
\bibitem[{Hsu et~al.(2012)Hsu, Weir, Truss, Garvey, Nicholson \&
  Halley}]{HSU20122385}
\bibinfo{author}{Hsu, Y.-C.}, \bibinfo{author}{Weir, M.~P.},
  \bibinfo{author}{Truss, R.~W.}, \bibinfo{author}{Garvey, C.~J.},
  \bibinfo{author}{Nicholson, T.~M.}, \& \bibinfo{author}{Halley, P.~J.}
  (\bibinfo{year}{2012}).
\newblock \bibinfo{title}{A fundamental study on photo-oxidative degradation of
  linear low density polyethylene films at embrittlement}.
\newblock {\it \bibinfo{journal}{Polymer}\/},  {\it \bibinfo{volume}{53}\/},
  \bibinfo{pages}{2385 -- 2393}.
\bibitem[{Hsueh et~al.(2020)Hsueh, Kim, Orski, Fairbrother, Jacobs, Perry,
  Hunston, White \& Sung}]{hsueh2020micro}
\bibinfo{author}{Hsueh, H.-C.}, \bibinfo{author}{Kim, J.~H.},
  \bibinfo{author}{Orski, S.}, \bibinfo{author}{Fairbrother, A.},
  \bibinfo{author}{Jacobs, D.}, \bibinfo{author}{Perry, L.},
  \bibinfo{author}{Hunston, D.}, \bibinfo{author}{White, C.}, \&
  \bibinfo{author}{Sung, L.} (\bibinfo{year}{2020}).
\newblock \bibinfo{title}{Micro and macroscopic mechanical behaviors of
  high-density polyethylene under uv irradiation and temperature}.
\newblock {\it \bibinfo{journal}{Polymer Degradation and Stability}\/},  {\it
  \bibinfo{volume}{174}\/}, \bibinfo{pages}{109098}.
\bibitem[{Johlitz et~al.(2014)Johlitz, Diercks \& Lion}]{johlitz2014thermo}
\bibinfo{author}{Johlitz, M.}, \bibinfo{author}{Diercks, N.}, \&
  \bibinfo{author}{Lion, A.} (\bibinfo{year}{2014}).
\newblock \bibinfo{title}{Thermo-oxidative ageing of elastomers: A modelling
  approach based on a finite strain theory}.
\newblock {\it \bibinfo{journal}{International Journal of Plasticity}\/},  {\it
  \bibinfo{volume}{63}\/}, \bibinfo{pages}{138--151}.
\bibitem[{Julienne et~al.(2019)Julienne, Lagarde \&
  Delorme}]{julienne2019influence}
\bibinfo{author}{Julienne, F.}, \bibinfo{author}{Lagarde, F.}, \&
  \bibinfo{author}{Delorme, N.} (\bibinfo{year}{2019}).
\newblock \bibinfo{title}{Influence of the crystalline structure on the
  fragmentation of weathered polyolefines}.
\newblock {\it \bibinfo{journal}{Polymer Degradation and Stability}\/},  {\it
  \bibinfo{volume}{170}\/}, \bibinfo{pages}{109012}.
\bibitem[{Kershaw \& Rochman(2015)}]{kershaw2015sources}
\bibinfo{author}{Kershaw, P.}, \& \bibinfo{author}{Rochman, C.}
  (\bibinfo{year}{2015}).
\newblock \bibinfo{title}{Sources, fate and effects of microplastics in the
  marine environment: part 2 of a global assessment}.
\newblock {\it \bibinfo{journal}{Reports and
  Studies-IMO/FAO/Unesco-IOC/WMO/IAEA/UN/UNEP Joint Group of Experts on the
  Scientific Aspects of Marine Environmental Protection (GESAMP) Eng No.
  93}\/}, .
\bibitem[{Lamnii et~al.(2021)Lamnii, Abdelaziz, Ayoub, Colin \&
  Maschke}]{lamnii2021experimental}
\bibinfo{author}{Lamnii, H.}, \bibinfo{author}{Abdelaziz, M.~N.},
  \bibinfo{author}{Ayoub, G.}, \bibinfo{author}{Colin, X.}, \&
  \bibinfo{author}{Maschke, U.} (\bibinfo{year}{2021}).
\newblock \bibinfo{title}{Experimental investigation and modeling attempt on
  the effects of ultraviolet aging on the fatigue behavior of an {LDPE}
  semi-crystalline polymer}.
\newblock {\it \bibinfo{journal}{International Journal of Fatigue}\/},  {\it
  \bibinfo{volume}{142}\/}, \bibinfo{pages}{105952}.
\bibitem[{Lee(1969)}]{lee1969elastic}
\bibinfo{author}{Lee, E.~H.} (\bibinfo{year}{1969}).
\newblock \bibinfo{title}{Elastic-plastic deformation at finite strains}.
\newblock {\it \bibinfo{journal}{Journal of Applied Mechanics}\/},  {\it
  \bibinfo{volume}{36}\/}, \bibinfo{pages}{1--6}.
\bibitem[{Makradi et~al.(2005)Makradi, Ahzi, Gregory \& Edie}]{makradi2005two}
\bibinfo{author}{Makradi, A.}, \bibinfo{author}{Ahzi, S.},
  \bibinfo{author}{Gregory, R.}, \& \bibinfo{author}{Edie, D.}
  (\bibinfo{year}{2005}).
\newblock \bibinfo{title}{A two-phase self-consistent model for the deformation
  and phase transformation behavior of polymers above the glass transition
  temperature: application to {PET}}.
\newblock {\it \bibinfo{journal}{International Journal of Plasticity}\/},  {\it
  \bibinfo{volume}{21}\/}, \bibinfo{pages}{741--758}.
\bibitem[{Peter et~al.(2006)Peter, Meyer \& Baschnagel}]{peter2006thickness}
\bibinfo{author}{Peter, S.}, \bibinfo{author}{Meyer, H.}, \&
  \bibinfo{author}{Baschnagel, J.} (\bibinfo{year}{2006}).
\newblock \bibinfo{title}{Thickness-dependent reduction of the glass-transition
  temperature in thin polymer films with a free surface}.
\newblock {\it \bibinfo{journal}{Journal of Polymer Science Part B: Polymer
  Physics}\/},  {\it \bibinfo{volume}{44}\/}, \bibinfo{pages}{2951--2967}.
\bibitem[{Rabek(1994)}]{rabek1994polymer}
\bibinfo{author}{Rabek, J.~F.} (\bibinfo{year}{1994}).
\newblock {\it \bibinfo{title}{Polymer photodegradation: mechanisms and
  experimental methods}\/}.
\newblock \bibinfo{publisher}{Springer Science \& Business Media}.
\bibitem[{Rabello \& White(1997)}]{rabello1997crystallization}
\bibinfo{author}{Rabello, M.}, \& \bibinfo{author}{White, J.}
  (\bibinfo{year}{1997}).
\newblock \bibinfo{title}{Crystallization and melting behaviour of
  photodegraded polypropylene—{I}. chemi-crystallization}.
\newblock {\it \bibinfo{journal}{Polymer}\/},  {\it \bibinfo{volume}{38}\/},
  \bibinfo{pages}{6379--6387}.
\bibitem[{Ranjan \& Goel(2019)}]{ranjan2019degradation}
\bibinfo{author}{Ranjan, V.~P.}, \& \bibinfo{author}{Goel, S.}
  (\bibinfo{year}{2019}).
\newblock \bibinfo{title}{Degradation of low-density polyethylene film exposed
  to uv radiation in four environments}.
\newblock {\it \bibinfo{journal}{Journal of Hazardous, Toxic, and Radioactive
  Waste}\/},  {\it \bibinfo{volume}{23}\/}, \bibinfo{pages}{04019015}.
\bibitem[{Rodriguez et~al.(2020)Rodriguez, Mansoor, Ayoub, Colin \&
  Benzerga}]{rodriguez2020effect}
\bibinfo{author}{Rodriguez, A.}, \bibinfo{author}{Mansoor, B.},
  \bibinfo{author}{Ayoub, G.}, \bibinfo{author}{Colin, X.}, \&
  \bibinfo{author}{Benzerga, A.} (\bibinfo{year}{2020}).
\newblock \bibinfo{title}{Effect of {UV}-aging on the mechanical and fracture
  behavior of low density polyethylene}.
\newblock {\it \bibinfo{journal}{Polymer Degradation and Stability}\/},  {\it
  \bibinfo{volume}{180}\/}, \bibinfo{pages}{109185}.
\bibitem[{Sauerbrey(1959)}]{sauerbrey1959verwendung}
\bibinfo{author}{Sauerbrey, G.} (\bibinfo{year}{1959}).
\newblock \bibinfo{title}{Verwendung von schwingquarzen zur w{\"a}gung
  d{\"u}nner schichten und zur mikrow{\"a}gung}.
\newblock {\it \bibinfo{journal}{Zeitschrift f{\"u}r physik}\/},  {\it
  \bibinfo{volume}{155}\/}, \bibinfo{pages}{206--222}.
\bibitem[{Shakiba et~al.(2016)Shakiba, Darabi \&
  Al-Rub}]{shakiba2016thermodynamic}
\bibinfo{author}{Shakiba, M.}, \bibinfo{author}{Darabi, M.~K.}, \&
  \bibinfo{author}{Al-Rub, R. K.~A.} (\bibinfo{year}{2016}).
\newblock \bibinfo{title}{A thermodynamic framework for constitutive modeling
  of coupled moisture-mechanical induced damage in partially saturated viscous
  porous media}.
\newblock {\it \bibinfo{journal}{Mechanics of Materials}\/},  {\it
  \bibinfo{volume}{96}\/}, \bibinfo{pages}{53--75}.
\bibitem[{Shakiba \& Najmeddine(2021)}]{shakiba2021physics}
\bibinfo{author}{Shakiba, M.}, \& \bibinfo{author}{Najmeddine, A.}
  (\bibinfo{year}{2021}).
\newblock \bibinfo{title}{Physics-based constitutive equation for
  thermo-chemical aging in elastomers based on crosslink density evolution}.
\newblock {\it \bibinfo{journal}{arXiv preprint arXiv:2104.09001}\/}, .
\bibitem[{Shlyapintokh(1983)}]{shlyapintokh1983synergistic}
\bibinfo{author}{Shlyapintokh, V.~Y.} (\bibinfo{year}{1983}).
\newblock \bibinfo{title}{Synergistic phenomena in polymer photostabilization}.
\newblock {\it \bibinfo{journal}{Pure and Applied Chemistry}\/},  {\it
  \bibinfo{volume}{55}\/}, \bibinfo{pages}{1661--1668}.
\bibitem[{Soares et~al.(2008)Soares, Moore~Jr \&
  Rajagopal}]{soares2008constitutive}
\bibinfo{author}{Soares, J.~S.}, \bibinfo{author}{Moore~Jr, J.~E.}, \&
  \bibinfo{author}{Rajagopal, K.~R.} (\bibinfo{year}{2008}).
\newblock \bibinfo{title}{Constitutive framework for biodegradable polymers
  with applications to biodegradable stents}.
\newblock {\it \bibinfo{journal}{Asaio Journal}\/},  {\it
  \bibinfo{volume}{54}\/}, \bibinfo{pages}{295--301}.
\bibitem[{Soares et~al.(2010)Soares, Rajagopal \&
  Moore}]{soares2010deformation}
\bibinfo{author}{Soares, J.~S.}, \bibinfo{author}{Rajagopal, K.~R.}, \&
  \bibinfo{author}{Moore, J.~E.} (\bibinfo{year}{2010}).
\newblock \bibinfo{title}{Deformation-induced hydrolysis of a degradable
  polymeric cylindrical annulus}.
\newblock {\it \bibinfo{journal}{Biomechanics and modeling in
  mechanobiology}\/},  {\it \bibinfo{volume}{9}\/}, \bibinfo{pages}{177--186}.
\bibitem[{Suresh et~al.(2011)Suresh, Maruthamuthu, Kannan \&
  Chandramohan}]{suresh2011mechanical}
\bibinfo{author}{Suresh, B.}, \bibinfo{author}{Maruthamuthu, S.},
  \bibinfo{author}{Kannan, M.}, \& \bibinfo{author}{Chandramohan, A.}
  (\bibinfo{year}{2011}).
\newblock \bibinfo{title}{Mechanical and surface properties of low-density
  polyethylene film modified by photo-oxidation}.
\newblock {\it \bibinfo{journal}{Polymer Journal}\/},  {\it
  \bibinfo{volume}{43}\/}, \bibinfo{pages}{398--406}.
\bibitem[{Tavares et~al.(2003)Tavares, Gulmine, Lepienski \&
  Akcelrud}]{tavares2003effect}
\bibinfo{author}{Tavares, A.~C.}, \bibinfo{author}{Gulmine, J.~V.},
  \bibinfo{author}{Lepienski, C.~M.}, \& \bibinfo{author}{Akcelrud, L.}
  (\bibinfo{year}{2003}).
\newblock \bibinfo{title}{The effect of accelerated aging on the surface
  mechanical properties of polyethylene}.
\newblock {\it \bibinfo{journal}{Polymer Degradation and Stability}\/},  {\it
  \bibinfo{volume}{81}\/}, \bibinfo{pages}{367--373}.
\bibitem[{Tervoort et~al.(1997)Tervoort, Smit, Brekelmans \&
  Govaert}]{tervoort1997constitutive}
\bibinfo{author}{Tervoort, T.}, \bibinfo{author}{Smit, R.},
  \bibinfo{author}{Brekelmans, W.}, \& \bibinfo{author}{Govaert, L.}
  (\bibinfo{year}{1997}).
\newblock \bibinfo{title}{A constitutive equation for the elasto-viscoplastic
  deformation of glassy polymers}.
\newblock {\it \bibinfo{journal}{Mechanics of Time-Dependent Materials}\/},
  {\it \bibinfo{volume}{1}\/}, \bibinfo{pages}{269--291}.
\bibitem[{Tireau et~al.(2009)Tireau, Van~Schoors, Benzarti \&
  Colin}]{tireau2009environmental}
\bibinfo{author}{Tireau, J.}, \bibinfo{author}{Van~Schoors, L.},
  \bibinfo{author}{Benzarti, K.}, \& \bibinfo{author}{Colin, X.}
  (\bibinfo{year}{2009}).
\newblock \bibinfo{title}{Environmental ageing of carbon black-filled
  polyethylene sheaths employed in civil engineering}.
\newblock {\it \bibinfo{journal}{J. Nanostruct. Polym. Nanocompos}\/},  {\it
  \bibinfo{volume}{5}\/}, \bibinfo{pages}{94--100}.
\bibitem[{Vieira et~al.(2011)Vieira, Marques, Guedes \&
  Tita}]{vieira2011material}
\bibinfo{author}{Vieira, A.}, \bibinfo{author}{Marques, A.},
  \bibinfo{author}{Guedes, R.}, \& \bibinfo{author}{Tita, V.}
  (\bibinfo{year}{2011}).
\newblock \bibinfo{title}{Material model proposal for biodegradable materials}.
\newblock {\it \bibinfo{journal}{Procedia Engineering}\/},  {\it
  \bibinfo{volume}{10}\/}, \bibinfo{pages}{1597--1602}.
\bibitem[{Vieira et~al.(2014)Vieira, Guedes \& Tita}]{vieira2014constitutive}
\bibinfo{author}{Vieira, A.~C.}, \bibinfo{author}{Guedes, R.~M.}, \&
  \bibinfo{author}{Tita, V.} (\bibinfo{year}{2014}).
\newblock \bibinfo{title}{Constitutive modeling of biodegradable polymers:
  Hydrolytic degradation and time-dependent behavior}.
\newblock {\it \bibinfo{journal}{International Journal of Solids and
  Structures}\/},  {\it \bibinfo{volume}{51}\/}, \bibinfo{pages}{1164--1174}.
\bibitem[{Wang et~al.(2004)Wang, Ge, Rafailovich, Sokolov, Zou, Ade,
  L{\"u}ning, Lustiger \& Maron}]{wang2004crystallization}
\bibinfo{author}{Wang, Y.}, \bibinfo{author}{Ge, S.},
  \bibinfo{author}{Rafailovich, M.}, \bibinfo{author}{Sokolov, J.},
  \bibinfo{author}{Zou, Y.}, \bibinfo{author}{Ade, H.},
  \bibinfo{author}{L{\"u}ning, J.}, \bibinfo{author}{Lustiger, A.}, \&
  \bibinfo{author}{Maron, G.} (\bibinfo{year}{2004}).
\newblock \bibinfo{title}{Crystallization in the thin and ultrathin films of
  poly (ethylene- vinyl acetate) and linear low-density polyethylene}.
\newblock {\it \bibinfo{journal}{Macromolecules}\/},  {\it
  \bibinfo{volume}{37}\/}, \bibinfo{pages}{3319--3327}.
\bibitem[{Wang et~al.(2010)Wang, Han, Pan \& Sinka}]{wang2010entropy}
\bibinfo{author}{Wang, Y.}, \bibinfo{author}{Han, X.}, \bibinfo{author}{Pan,
  J.}, \& \bibinfo{author}{Sinka, C.} (\bibinfo{year}{2010}).
\newblock \bibinfo{title}{An entropy spring model for the young’s modulus
  change of biodegradable polymers during biodegradation}.
\newblock {\it \bibinfo{journal}{Journal of the Mechanical Behavior of
  Biomedical Materials}\/},  {\it \bibinfo{volume}{3}\/},
  \bibinfo{pages}{14--21}.
\bibitem[{Xiao \& Nguyen(2016)}]{XIAO201670}
\bibinfo{author}{Xiao, R.}, \& \bibinfo{author}{Nguyen, T.~D.}
  (\bibinfo{year}{2016}).
\newblock \bibinfo{title}{A thermodynamic modeling approach for dynamic
  softening in glassy amorphous polymers}.
\newblock {\it \bibinfo{journal}{Extreme Mechanics Letters}\/},  {\it
  \bibinfo{volume}{8}\/}, \bibinfo{pages}{70--77}.
\bibitem[{Yousif \& Haddad(2013)}]{yousif2013photodegradation}
\bibinfo{author}{Yousif, E.}, \& \bibinfo{author}{Haddad, R.}
  (\bibinfo{year}{2013}).
\newblock \bibinfo{title}{Photodegradation and photostabilization of polymers,
  especially polystyrene}.
\newblock {\it \bibinfo{journal}{SpringerPlus}\/},  {\it
  \bibinfo{volume}{2}\/}, \bibinfo{pages}{1--32}.
\bibitem[{Zhao \& Zikry(2017)}]{zhao2017oxidation}
\bibinfo{author}{Zhao, B.}, \& \bibinfo{author}{Zikry, M.}
  (\bibinfo{year}{2017}).
\newblock \bibinfo{title}{Oxidation-induced failure in semi-crystalline organic
  thin films}.
\newblock {\it \bibinfo{journal}{International Journal of Solids and
  Structures}\/},  {\it \bibinfo{volume}{109}\/}, \bibinfo{pages}{72--83}.
\bibitem[{Zhao et~al.(2020)Zhao, Lei, Zhang, Chen \& Fang}]{ZHAO2020100826}
\bibinfo{author}{Zhao, Z.}, \bibinfo{author}{Lei, M.}, \bibinfo{author}{Zhang,
  Q.}, \bibinfo{author}{Chen, H.-S.}, \& \bibinfo{author}{Fang, D.}
  (\bibinfo{year}{2020}).
\newblock \bibinfo{title}{A general model for the temperature-dependent
  deformation and tensile failure of photo-cured polymers}.
\newblock {\it \bibinfo{journal}{Extreme Mechanics Letters}\/},  {\it
  \bibinfo{volume}{39}\/}, \bibinfo{pages}{100826}.

\end{thebibliography}

\end{document}